\documentclass{jpp}
\usepackage{graphicx}
\usepackage{epstopdf,epsfig}
\usepackage{amsmath,amssymb}
\usepackage{natbib}

\newcommand{\ie}{\emph{i.e., }}
\newcommand{\reff}[1]{(\ref{#1})}
\newcommand{\eref}[1]{Eq.\reff{#1}}
\newcommand{\erefs}[1]{Eqs.\reff{#1}}
\newcommand{\figref}[1]{FIG.\ref{#1}}
\newcommand{\params}{\Theta}
\newcommand{\deltat}{\Delta \tau}

\newcommand{\np}{n_p}
\newcommand{\nb}{n_B}

\newcommand{\Nu}{N_1}
\newcommand{\Nd}{N_2}
\newcommand{\Nn}{N_n}
\newcommand{\su}{\sigma_1}
\newcommand{\sd}{\sigma_2}
\newcommand{\sn}{\sigma_n}

\newcommand{\sumiu}{\sum_{i=1}^{\Nu}}
\newcommand{\sumid}{\sum_{i=1}^{\Nd}}
\newcommand{\sumin}{\sum_{i=1}^{\Nn}}
\newcommand{\kj}{k_j}
\newcommand{\ku}{k_1}

\newcommand{\kd}{k_2}

\newcommand{\lj}{\ell_j}
\newcommand{\lu}{\ell_1}

\newcommand{\ld}{\ell_2}
\newcommand{\omp}{\omega_p}

\newcommand{\omjt}{\omega}

\newcommand{\vu}{v_1}
\newcommand{\vd}{v_2}
\newcommand{\vn}{v_n}

\newcommand{\phij}{\varphi_j}

\newcommand{\phija}{\phi_j}
\newcommand{\phiua}{\phi_1}
\newcommand{\phida}{\phi_2}

\newcommand{\eps}{\epsilon_p}

\newcommand{\epsp}{\epsilon^{\scriptscriptstyle{\prime}}}

\newcommand{\xbui}{\bar{\xi}_{1i}}
\newcommand{\xbdi}{\bar{\xi}_{2i}}

\newcommand{\zui}{\zeta_{1i}}
\newcommand{\zdi}{\zeta_{2i}}
\newcommand{\zni}{\zeta_{ni}}
\newcommand{\zbui}{\bar{\zeta}_{1i}}

\newcommand{\zbni}{\bar{\zeta}_{ni}}
\newcommand{\betauj}{\beta_{1j}}

\newcommand{\betanj}{\beta_{nj}}
\newcommand{\etab}{\bar{\eta}}
\newcommand{\phijaz}{\phi_{j}^{0}}

\newcommand{\ziz}{\bar{\zeta}_{ai}^{0}}
\newcommand{\zipz}{\bar{\zeta}_{ai}^{\prime 0}}

\shorttitle{Nonlinear physics of the beam-plasma instability}
\shortauthor{N. Carlevaro, M.V. Falessi, G. Montani, F. Zonca}

\title{Nonlinear physics and energetic particle transport features of the beam-plasma instability}

\author{Nakia Carlevaro\aff{1}\corresp{\email{nakia.carlevaro@gmail.com}},
        Matteo V. Falessi\aff{2},\\
        Giovanni Montani\aff{1,3} \and
        Fulvio Zonca\aff{1}}

\affiliation{
\aff{1}ENEA for EUROfusion,\\
       Via E. Fermi, 45 (00044) Frascati (Roma), Italy.
\aff{2}Department of Physics, ``RomaTre'' University,\\
       Via della Vasca Navale, 84 (00146) Roma, Italy.
\aff{3}Department of Physics, ``Sapienza'' University of Rome,\\
       P.le Aldo Moro, 5 (00185) Roma, Italy.}

\begin{document}

\maketitle

\begin{abstract}
In this paper, we study transport features of a one-dimensional beam-plasma system in the presence of multiple resonances. As a model description of the general problem of a warm energetic particle beam, we assume $n$ cold supra-thermal beams and investigate the self-consistent evolution in the presence of the complete spectrum of nearly degenerate Langmuir modes. A qualitative transport estimation is obtained by computing the Lagrangian Coherent Structures of the system on given temporal scales. This leads to the splitting of the phase space into regions where the local transport processes are relatively faster. The general theoretical framework is applied to the case of the nonlinear dynamics of two cold beams, for which numerical simulation results are illustrated and analyzed.
\end{abstract}

\section{Introduction}
Nonlinear interplay of energetic particles (EPs) with Alfv\'enic fluctuations, such as Alfv\'en eigenmodes (AEs), EP modes and drift Alfv\'en waves \citep{ZCrmp,CZ07,CZ13,Fa07}, and their consequences for fluctuation induced cross-field transport constitute basic phenomena for thermonuclear plasmas \citep{Fa07,H08,BS11,La13,VB13,WB11,ML13,ML12,GPT14,ZC15ppcf,ZC15njp,Pi15}. The relevance of the beam-plasma instability problem as paradigm for EP transport due to AEs was proposed in \citep{BB90a,BB90b,BB90c}. In fact, the bump-on-tail paradigm can mimic the nonlinear interaction of supra-thermal particles with Alfv\'enic fluctuations, provided the system is sufficiently close to marginal stability \citep{CZ07,CZ13,BB90c,BS11}. In particular, it is possible to determine a correspondence between velocity space transport in the one-dimensional (1D) beam-plasma instability problem and radial transport of fast ions in the presence of resonant AEs, at least as far as their spatial structure does not play a significant role. More precisely, such correspondence has rigorous foundation as long as the nonlinear resonant particle displacement is small compared to the characteristic perpendicular fluctuation scale  \citep{ZCrmp,ZC15ppcf,ZC15njp,WB12pre,BW14pop}.

The beam-plasma instability was analyzed in the pioneering works of O'Neil and collaborators \citep{OWM71,OM68} [see also \citep{Sh63,OL70,SS71,Th71,L72,Ma72}]. It was analyzed for a nearly monochromatic fast electron beam injected into a 1D plasma, which, in turn, is treated as a linear dielectric medium supporting longitudinal Langmuir waves \citep{LP81}. It was demonstrated that Langmuir fluctuations can be resonantly excited by EPs and become unstable. The original approach of \citep{OWM71} can be cast in a rigorous Hamiltonian formulation \citep{MK78,TMM94,AEE98}; and the resulting scheme is formally equivalent to that of the Free Electron Laser dynamics \citep{AJFR98}. An alternative approach was given in \citep{AK66}, where the nonlinear Vlasov-Poisson initial value problem was solved for the case of nearly degenerate Langmuir and beam modes and an analogue of the Dyson equation was derived [see \citep{ZCrmp,ZC15ppcf,ZC15njp} for an in-depth discussion of this approach]. An example of the beam-plasma instability in the presence of more than a single resonant mode was analyzed by \citep{ZK08}, emphasizing the link with the so-called sideband instability, and addressing a simple Hamiltonian formulation of the problem. Meanwhile, \citep{EHM03} also include the effect of a magnetic field in the multi-wave model, but the analysis refers to a single longitudinal resonant electrostatic mode. Pioneering works on these thematics can be found in \citep{DeN75} and in \citep{LKK75}, where the 1D beam-plasma interaction with higher harmonics has been studied.

The beam-plasma problem is characterized by typical velocity and temporal scales. Nonlinear saturation of the beam-plasma instability occurs over a $\sim\gamma_L^{-1}$ characteristic time, with $\gamma_L$ the linear instability growth rate \citep{OM68}; while beam heating typically takes place at a rate $\sim\omega_B\propto|ek^{2}\varphi/m_e|^{1/2}$, \ie the wave-particle trapping frequency in the fluctuating scalar potential $\varphi$ with wave number $k$ (here, $m_e$ and $e$ denote the mass and charge of electrons, respectively) and involves resonant particles in an interval $\Delta v\sim\omega_B/k\sim(\omega_B/\omp) v_B$ about the velocity resonance condition $v=v_B$ (where $v_B$ is the EP drifting beam velocity). The saturation level of the beam-plasma instability can be estimated as $\omega_B\sim3\gamma_L$ \citep{OWM71,L72}, while the flattening of the beam distribution function can take much longer times than $\sim\omega_B^{-1}$ and involves the formation of structures in the wave-particle trapping region \citep{TMM94}. It has been recently shown \citep{CFGGMP14} that the nonlinear evolution of the beam-plasma system is characterized by quasi-stationary states which can be accurately predicted by the maximum entropy principle proposed by Lynden-Bell \citep{AEE98}.

In the present work, we formulate the general problem of $n$ cold beams self-consistently evolving in the presence of $m\geq n$ unstable modes degenerate with the Langmuir potentials at the plasma frequency $\omp$, \ie $\omega=\omp$, which ensures that the dielectric function of the cold background plasma is nearly vanishing. Among these waves, we can select $n$ modes which are resonantly driven by the beams [obeying to the O'Neil selection rule $k=\omp/v_B$ \citep{OM68,OWM71}] and $m-n$ quasi-degenerate ones associated with nonlinear wave-particle and/or mode-mode coupling in the case of (almost) overlapping resonances. It is worth noting that the dominant quasi-degenerate modes can also be dealt with as slowly evolving Langmuir waves when the spectral density is sufficiently large. In particular, we will show in Sec.~\ref{simsnlth} that nonlinear excitation of quasi-degenerate sidebands can be important for the case of two interacting beams.

The physical motivation for the proposed model relies on the fact that, due to the intrinsic beam heating involved in the nonlinear beam-plasma dynamics \citep{Sh63}, transport is not qualitatively different for cold and warm beams as long as the initial beam width in velocity space is smaller than $\Delta v$. Thus, transport in the presence of overlapping resonances may be studied as the evolution of initially cold beams as a consequence of the corresponding nearly degenerate beam (Langmuir) instabilities. This statement justifies the representation of the bump-on-tail problem in terms of $n$ overlapping resonances and has the valuable feature of maintaining direct control on spectral density and intensity, which are the properties that ultimately determine EP transport.

The proposed theoretical model can properly describe the case of isolated resonances and $m=n$ uncoupled nonlinear oscillators (dominated by wave-particle trapping), as well as the ``quasi-linear'' diffusion regime typical of a ``broad fluctuation spectrum'' \citep{AK66}. In fact, the nonlinear equation system we propose, in the limit of very large $n$, is expected to be physically isomorphic to quasi-linear models of EP interaction with AEs. In this respect, it is important to mention that a line-broadened quasi-linear approach has been proposed by \citep{BB95b,BB96b} for computing EP transport by means of a diffusion equation, which could address not only overlapping modes but also the broadening of the resonant spectrum for isolated instabilities in the case of multiple AEs. This model has been extended and compared with experimental observations \citep{GG12} and with numerical solutions of the bump-on-tail paradigm \citep{GBG14}. Contrary to this approach, which cannot account for frequency chirping phenomena \citep{GBG14}, the present $n$-beam model aims at characterizing particle transport on time scale comparable to the instability growth-rate and, thus, it can address a broad range of underlying complex phenomena \citep{ZCrmp,ZC15ppcf,ZC15njp}.

As an illustration of the onset of particle transport phenomena in the phase space, the simulation discussed in this work are specialized to two cold beams and corresponding instabilities. Clearly, when the two beams have sufficiently different initial velocities, we observe the growth and saturation of two independent modes and transport processes are practically suppressed. Meanwhile, when the two velocities become closer, a regime exists when the slower beam significantly interacts with the resonant mode excited by the faster beam, and an ${\cal O}(1)$ transfer of energy from the latter mode to the former takes place. Finally, when the velocity ratio exceeds a critical threshold consistent with the Chirikov criterion \citep{L72,Ch79}, significant phase-space mixing takes place between the two beam populations.

As we are interested in the details of the the mixing process between supra-thermal particles belonging to different beams in the presence of unstable fluctuations, we will present an analysis using a Lagrangian approach in order to quantitatively define the transport barriers of the system by means of the so-called Lagrangian Coherent Structures (LCSs), introduced for the first time by \citep{PH13} and defined in terms of the features of the Finite Time Lyapunov Exponent (FTLE) fields \citep{MPJPP}. For practical reasons, particle transport is described as the evolution of single passive tracers (test particles) in a time-dependent potential generated at each instant by the self-consistent charge distribution. This approach, which is equivalent to the original problem, leads to important conceptual simplifications if we are interested in studying the onset of an instability that naturally selects only some Fourier components. The advection of passive tracers in a non-periodic flow has been studied by \citep{SLM05,H11}, leading to the definition of LCSs. These structures, which are 1D curves in our system, are finite-time transport barriers for the tracers and, therefore, play the role of the last periodic surface of the Poincar\'e map introduced in \citep{SLM05,H11}. Thus, they divide tracers into two classes with a qualitatively different evolution. This way of characterizing transport gives a criterion for the transition to local chaotic behavior, \ie for assessing when nonlinear dynamics causes neighboring cold beams to merge.

The paper is organized as follows. In Sec.\ref{nonlinearth}, the nonlinear treatment of the beam-plasma system of O'Neil is generalized in the presence of $n$ distinct beams and $m$ nearly degenerate Langmuir modes. In Sec.\ref{simsnlth}, the theoretical model is then specialized for the two beam case and, in the first subsection, for the corresponding two most unstable modes. Numerical simulations are presented first by analyzing temporal evolutions of scalar potentials and then by discussing the phase-space dynamics. Mixing phenomena between different populations are qualitatively analyzed as a function of the ratio between initial beam velocities. An analysis of quasi-degenerate modes including nonlinear sideband excitation is discussed in the second subsection. In Sec.\ref{ftlesection}, transport processes are qualitatively studied by means of the LCS technique. Phase-space mixing is found to be enhanced at the intersections between LCSs belonging to different beam populations. This analysis allows us to define a critical initial velocity ratio of beams as the smallest value leading to beam mixing. Concluding remarks are given in Sec.\ref{conclu}.

\section{Nonlinear evolution of \emph{n} resonant beams}\label{nonlinearth}

In the original stability analysis of the beam-plasma system \citep{OM68}, it was shown how fluctuations can be resonantly excited by an injected supra-thermal beam (of velocity $v_B$). Introducing the plasma frequency $\omp\equiv\sqrt{4\pi\np e^2/m_e}$ ($\np$ being the plasma density), two types of modes were shown to exist as solutions of the dispersion relation: \emph{(i)} the stable Langmuir wave with $\omega=\omp$, modified by the presence of EPs; and \emph{(ii)} the beam mode, with both stable and unstable branches at frequency $\omega\simeq k v_B$, and maximum growth rate when $\omega\simeq\omp$ (here, $k$ represents the wave number of the unstable longitudinal mode). Thus, most unstable beam modes are excited when they are ``nearly degenerate'' with Langmuir waves, \ie $k\simeq\omp/v_B$. Meanwhile, the nonlinear evolution takes place in two stages \citep{OWM71,Sh63}: first, beam-plasma interactions heat the beam as EPs begin sloshing back and forth in the potential well of the wave; second, the beam distribution function is eventually flattened in velocity space by nonlinear interactions. Thus, the process leading to saturation of the beam-plasma instability is due to wave trapping of resonant particles, whose transport in velocity space is connected with energy exchange between fast electrons and the electric field.

In this Section, we derive the governing equations for the general system of $n$ cold beams and $m\geq n$ self-consistently coupled modes \citep{OWM71}. It is important to stress that all considered $m$ modes are assumed to be nearly degenerate Langmuir waves with close to vanishing dielectric function, while resonance conditions are not imposed \emph{a priori}. The background plasma is a cold homogeneous linear dielectric medium with constant particle density $\np$. The dynamics of fast electrons is determined by the Newton's law. Equations of motion are closed by the Poisson equation which provides the time evolution for each of the nearly degenerate Langmuir waves considered in the model.

In a 1D model, the motion along the $x$ direction of period $L$ is labeled by $x_1$, $x_2$, ..., $x_n$ for each beam, respectively. Decomposing the single beams in $\Nu$, $\Nd$, $...$, $\Nn$ charge sheets located at $x_{1i}$, $x_{2i}$, ..., $x_{ni}$, the system can be discretized using the following total charge density
\begin{align}
\rho_B(x,t)=-e\,\nb\Big(\frac{L\su}{\Nu}\sumiu\delta(x-x_{1i})
+...
+\frac{L\sn}{\Nn}\sumin\delta(x-x_{ni})\Big)\;,
\end{align}
where $\nb$ is the total supra-thermal particle number density while $\su\nb$, $...$, $\sn\nb$ are the number densities of each beam. The Langmuir wave scalar potential can be expressed using the Fourier representation as
\begin{align}
\varphi(x,t)=\sum_{j=1}^{m}\big(\phij(\kj,t)e^{i\kj x-i\omp t}+c.c.\big)\;,
\end{align}
where $\phij(\kj,t)$ are assumed to be slowly varying fields and we have explicitly indicated the time dependence relative to the rapid oscillations associated to the plasma frequency (the c.c. denotes the complex conjugate). Thus, the $\phij$'s represent the effective nonlinear evolution of the potentials.

The trajectories of charge sheets are derived from the motion equation $\ddot{x}=e\nabla\varphi/m_e$ as
\begin{subequations}\label{forcex}
\begin{align}
\ddot{x}_{1i}=\frac{ie}{m}\sum_{j=1}^{m}\;\kj\phij e^{i\kj x_{1i}-i\omp t}+c.c.\;,\\
\vdots\nonumber\\
\ddot{x}_{ni}=\frac{ie}{m}\sum_{j=1}^{m}\;\kj\phij e^{i\kj x_{ni}-i\omp t}+c.c.\;.
\end{align}
\end{subequations}
Meanwhile, the evolution of the fields $\phij$ is determined by the Poisson's equation written in the Fourier space for fixed $\kj$ as
\begin{align}\label{poissonk}
\kj^{2}\eps\phij e^{-i\omp t}=4\pi\,\rho_{Bj}(\kj,t)\;,
\end{align}
where $\eps$ is the dielectric function (we remind that for a cold plasma $\eps=1-\omp^2/\omega^2$) and $\rho_{Bj}(\kj,t)$ is the Fourier transformed density
\begin{align}
\rho_{Bj}(\kj,t)=\frac{1}{L}\int_0^L\!\!\!\!dx\;\rho_B(x,t)e^{-i\kj x}=-e\,\nb\Big(
\frac{\su}{\Nu}\sumiu e^{-i\kj x_{1i}}
+...
+\frac{\sn}{\Nn}\sumin e^{-i\kj x_{ni}}\Big)\;.
\end{align}

For a cold plasma and $j=1,\,2,\,...,\,m$ nearly degenerate Langmuir waves, the dielectric function is nearly vanishing and can thus be expanded near $(\omp,\kj)$ as
\begin{align}
\eps(\omega(t),\kj)=\eps(\omp,\kj)+(\partial\eps/\partial\omega)\Big|_{\omp,\kj}\,(\omega_j-\omp)\equiv\epsp\,(\omjt_j-\omp)\;,
\end{align}
where the time-dependent $\omjt_j$ account for the nonlinear frequency shift with respect to $\omp$. Without loss of generality, the fields $\varphi_j$ can be now expressed as
\begin{align}\label{oihjpiuhpiuh}
\phij(t)\propto\exp(-i\int_0^t d\bar{t} (\omjt_j(\bar{t})-\omp))\;,
\end{align}
thus obtaining $\dot{\varphi}_j=-i(\omjt_j-\omp)\phij$; and \eref{poissonk} finally writes
\begin{align}\label{eqsphi}
\dot{\varphi}_j=i\,\frac{4\pi e \nb}{\kj^{2}\,\epsp}\;\Big(
\frac{\su}{\Nu}\sumiu e^{-i\kj x_{1i}+i\omp t}
+...
+\frac{\sn}{\Nn}\sumin e^{-i\kj x_{1i}+i\omp t}\Big)\;.
\end{align}
The equations of motion governing the beam-plasma system are \erefs{forcex} and \reff{eqsphi}.

Let us now analyze the effective nonlinear shift $\zeta$ of charge sheets due to the beam-wave interaction by decomposing the motion as
\begin{align}\label{xtoxi}
x_{1i}=\vu t+\zui(t)\;,\quad
x_{2i}=\vd t+\zdi(t)\;,\quad
...\;,\quad
x_{ni}=\vn t+\zni(t)\;.
\end{align}
Performing this coordinate transformation and introducing the scaled variables (we recall that, for a cold plasma, $\epsp=2/\omp$)
\begin{align}\label{adimvar}
\kj=2\pi\lj/L\;,\quad
\zbni=2\pi\zni/L\;,\quad
\tau=t\omp\etab\;,\quad
\phija=[\phij\,e\kj^2]/[m\etab^{2}\omp^2]\;,
\end{align}
with $\etab=(\nb/2\np)^{1/3}$ and the frequency mismatches
\begin{align}
\betauj=[\kj\vu-\omp]/\omp\etab\;,\quad
...\;,\quad
\betanj=[\kj\vn-\omp]/\omp\etab\;,
\end{align}
the set of equations describing the effective nonlinear dynamics of $n$ cold beams interacting with $m$ Langmuir modes reads now (the prime denotes the $\tau$ derivative)
\begin{subequations}\label{generalsystem}
\begin{align}
\zbui''&=i\;\sum_{j=1}^{m}\;\lj^{-1}\;\phija\;e^{i\lj\zbui+i\betauj\tau}+c.c.\;,\\
&\phantom{\zbui''=i\;\sum_{j=1}^{m}\;\lj^{-1}\;\phija\;e^{i\lj\zbui+i\betauj\tau}+cc}\vdots\nonumber\\
\zbni''&=i\;\sum_{j=1}^{m}\;\lj^{-1}\;\phija\;e^{i\lj\zbni+i\betanj\tau}+c.c.\;,\\
\phija'&=i\Big(\frac{\su}{\Nu}\sumiu e^{-i\lj\zbui-i\betauj\tau}
+...
+\frac{\sn}{\Nn}\sumin e^{-i\lj\zbni-i\betanj\tau}\Big)\;.\label{poieqadi}
\end{align}
\end{subequations}
The frequency mismatch terms are crucial (as in the single wave model) for defining which mode is growing fastest. The leading modes correspond to the ones which minimize the $\beta$'s implying that the most unstable modes associated to a given beam must obey the O'Neil selection rule $k_j=v_j/\omp$. We stress how this dynamical system is isomorphic with respect to the model commonly adopted in a class of studies of wave-particle interactions and rigorously derived in \citep{EEbook} (see also \citep{AEE98,AEFR06,FE00,Farina04,Farina94}). For an example of the mapping between the beam-plasma instability and the Free-Electron Laser model see \citep{Duccio05}.

The conservation laws of the system above can be derived by extension of the method outlined by \citep{OWM71}. The total momentum conservation reads now
\begin{align}\label{totmomcons}
\sum_{j=1}^{m}\frac{|\phija|^2}{\lj}+\frac{\su}{\Nu}\sumiu \zbui'+...+\frac{\sn}{\Nn}\sumin \zbni'=\mathcal{K}_M\;,
\end{align}
with constant $\mathcal{K}_M$. As already stated in \eref{oihjpiuhpiuh}, the potentials obey to the equation $\dot{\varphi}_j=-i(\omjt_j-\omp)\phij$, which can be simply rewritten as $\phija'=-i\Omega_j\phija$, where $\Omega_j\equiv(\omega_j-\omp)/(\omp\etab)=\delta\omega_j/\omp\etab$. Using these expressions, the total energy conservation can be cast in the following form
\begin{align}
\sum_{j=1}^{m}&\frac{|\phija|^2}{\etab\lj^2}+2\sum_{j=1}^{m}\Omega_j^{Re}\frac{|\phija|^2}{\lj^2}+\nonumber\\
&+\frac{1}{2}\Big[\frac{\su}{\Nu}\sumiu(\zbui'+2\pi\vu/L\etab\omp)^2+...
+\frac{\sn}{\Nn}\sumin(\zbni'+2\pi v_n/L\etab\omp)^2\Big]=\mathcal{K}_E\;,
\label{totenecons}
\end{align}
with constant $\mathcal{K}_E$. These conservation laws are directly connected with the Hamiltonian representation of the system. Thus the present approach based on the motion of $N$ particles in the self-consistent Langmuir potentials is more general and reduces to the Vlasov-Poisson description in the mean field limit of the Hamiltonian formulation treated \emph{a la} Klimontovich \citep{Klim}. The Vlasov-Maxwell equations are exactly recovered taking the continuum or mush limit, \ie artificially increasing the number of particles and keeping the charge density constant. A discussion of this paradigm in view of the Lynden-Bell treatment of the distribution function for the beam-plasma model is presented in \citep{CFGGMP14,AJFR98}. The significant advantage in adopting the present discrete model relies on the simplicity of the governing equations for particle and field self-consistent evolution. Furthermore, the present scheme allows direct control of the spectral density (\ie the separation in velocity space of the beams) and on the spectral intensity (\ie the values of the $\sigma$'s) of the initial beam-plasma system.

As to the comparison of the present analysis with respect to the Hamiltonian approach presented in \citep{EEbook,EE08} [see also \citep{ZK08,KV14,VK12} and references therein], it is worth noting that they are closely related except for their analysis of an assigned warm beam in contrast to our approach, based on $n$ cold beams. Due to the analysis in \citep{L72}, the intrinsic beam heating by nonlinear wave-particle interactions extends up to a velocity width of the order $\Delta v\simeq \omega_B/k\simeq\gamma_L/k\simeq 3\sqrt{3}\etab\omp/2k$. When multiple resonant beams overlap, the formation of a continuous particle distribution function is simply due to the velocity spread associated to the beam heating. Thus, addressing the hot beam dynamics by $n$ cold overlapping resonant beams is a reasonable and reliable simplification of the nonlinear beam plasma dynamics.

Let us now derive the linear dispersion relation [see \citep{OM68} for the analysis of the linear beam-plasma instability]. As first step, we assume $\Omega_j$ independent of $\tau$, leading to the expression $\phija=\phijaz\,e^{-i\Omega_j\tau}$. The equation of motion can then be solved using linear perturbation theory by expanding $\bar{\zeta}_{ai}=\ziz+\zipz\;\tau+\mathcal{O}(\tau^{2})$, where the subscript denotes the ``a'' beam. The first-order integration yields
\begin{align}\label{xiexp}
\bar{\zeta}_{ai}=\ziz+\zipz\;\tau-i\sum_{j=1}^{m}
\frac{\lj^{-1}\phija\;\exp[i\lj\ziz+i\lj\zipz\tau+i\beta_{aj}\tau]}{(\Omega_j-\beta_{aj}-\lj\zipz)^{2}}+c.c.\;.
\end{align}
The Poisson equation \reff{poieqadi} can be integrated by substituting the expression above and by considering uniform initial conditions for the nonlinear shift such that $\sum_ie^{i\lj\ziz}=0$. Expanding the exponential in the small $\phija$ limit and neglecting $\mathcal{O}(\phi^2)$, we finally obtain
\begin{align}
\phija'=-i\Omega_j\phija=\sum_a\Big[-i\frac{\sigma_a}{N_a}\sum_{i=1}^{N_a}\frac{\phija}{(\Omega_j-\beta_{aj}-\lj\zipz)^{2}}\Big]\;.
\end{align}
According to \eref{xtoxi}, we consider $\zipz=0$; thus, the equation above reduces to the following dispersion relation
\begin{align}\label{disrelnonlin}
\Omega_j=\frac{\su}{(\Omega_j-\beta_{1j})^{2}}+\frac{\sd}{(\Omega_j-\beta_{2j})^{2}}+...+\frac{\sn}{(\Omega_j-\beta_{nj})^{2}}\;,
\end{align}
where the presence of the $j$-index represents the corresponding mode for which the relation is evaluated; and we recall that $\beta_{jj}=0$.

To conclude this Section, since the proposed model (as discussed above) incorporates the bump-on-tail paradigm, let us now briefly discuss some features and issues of this problem. The bump-on-tail paradigm and its generalization in the presence of dissipation and collisions \citep{BB93,BB95a,BB96a,BB97a,BB97b,WB98} was extensively studied in \citep{LI09,LI10,LI12}, along with an in-depth discussion on its applicability limits. In these works, the 1D model was considered as a dimensional reduction of the 3D AE problem near an isolated resonance. Recently, it was also demonstrated that holes and clumps may be generated with any (small) amount of background dissipation, provided that a phase-space plateau is formed by phase mixing and dissipative damping of an unstable kinetic resonance \citep{LN14}. More precisely, in this case, holes and clumps are negative energy waves that grow because of background dissipation. The qualitative scenario of onset of stochastic transport in the presence of multiple resonances within the 1D bump-on-tail paradigm has been recently reviewed by \citep{BS11,Br11} and the implications of quasi-linear diffusion in the presence of many modes have been discussed by \citep{LB12}. In that work, it has been argued that extended flattening of particle distribution in the phase space can be more important than quasi-linear diffusion in the presence of many modes. The main limitation of these analyses is the adiabatic \emph{ansatz} for nonlinear dynamics, \ie the assumption that the characteristic nonlinear time $\tau_{NL}$ is much greater than $\omega_B^{-1}$ not only for quasi-linear diffusion but also for the evolution of chirping holes and clumps. Removing  this limitation is one of the main motivation of the analysis proposed in the present work. Meanwhile, the conjecture that long-range frequency sweeping events can be mapped to corresponding AE behaviors in fusion plasmas is questionable, due to the lack of mode structures in the theoretical description \citep{ZCrmp}. Finally, we note that the Vlasov-Poisson system (on which the bump-on-tail paradigm is based) has been recently studied by numerical simulations based on Eulerian \citep{Sho10} as well as Lagrangian codes \citep{PIC3D}.

\section{Simulations of the two resonant beam model}\label{simsnlth}
Let us now specialize the general problem described above to two supra-thermal beams. We first discuss the case when only the corresponding two most unstable waves are included in the dynamics; and we then address the problem of the presence of additional quasi-degenerate modes. It is worth noting that a similar problem was addressed by \citep{ZK08} in the framework of a Hamiltonian approach as in \citep{EEbook}. However, that work includes an external magnetic field in the beam-plasma dynamics; resulting in a 3D geometry of the problem. Since the dispersion relation of the considered case contains additional degrees of freedom (\emph{e.g.}, transverse wave numbers), a direct comparison with the present analysis is limited, especially in the respect of considering two distinct resonant velocities as below.

\subsection{Most unstable modes}\label{twomostmode}
The two most unstable modes are characterized by the resonance conditions $\ku=\omp/\vu$ and $\kd=\omp/\vd$. In what follows, we define $\params=\vd/\vu$ (thus implying $\ld\params=\lu$) and, since the system is specular with respect to the value $\params=1$, for the sake of simplicity, we report here only the results for $\params<1$. In order to properly represent the phase-space dynamics, we move to the average beam speed reference frame. The dynamics of the two beam sheets is thus represented as
\begin{align}
x_{1i}=\vu t+\zeta_{1i}\equiv t(\vu+\vd)/2+\xi_{1i}(t)\;,\quad
x_{2i}=\vd t+\zeta_{2i}\equiv t(\vu+\vd)/2+\xi_{2i}(t)\;,
\end{align}
where it worth noting that the time evolution is formally the same for the different representations of beam nonlinear shifts but with modified initial conditions. Introducing the frequency mismatch as
\begin{align}\label{betabar}
\beta_j=\big[\kj\,(\vu+\vd)/2-\omp\big]/\omp\etab\;,
\end{align}
and using the scaled variables \reff{adimvar} (here $\bar{\xi}_{ai}=2\pi\xi_{ai}/L$), the set of equations describing the nonlinear dynamics are
\begin{subequations}\label{mainsys}
\begin{align}
\xbui''&=i\,\big(\lu^{-1}\;\phiua\;e^{i\lu\xbui+i\beta_1 \tau}+\ld^{-1}\;\phida\;e^{i\ld\xbui+i\beta_2 \tau}\big)+c.c.\;,\\
\xbdi''&=i\,\big(\lu^{-1}\;\phiua\;e^{i\lu\xbdi+i\beta_1 \tau}+\ld^{-1}\;\phida\;e^{i\ld\xbdi+i\beta_2 \tau}\big)+c.c.\;,\\
\phiua'&=ie^{-i\beta_{1}\tau}\Big(\frac{\su}{\Nu}\sumiu e^{-i\lu\xbui}+\frac{\sd}{\Nd}\sumid e^{-i\lu\xbdi}\Big)\;,\\
\phida'&=ie^{-i\beta_{2}\tau}\Big(\frac{\su}{\Nu}\sumiu e^{-i\ld\xbui}+\frac{\sd}{\Nd}\sumid e^{-i\ld\xbdi}\Big)\;.
\end{align}
\end{subequations}

\begin{figure}
\begin{center}
\includegraphics[width=.35\textwidth,clip]{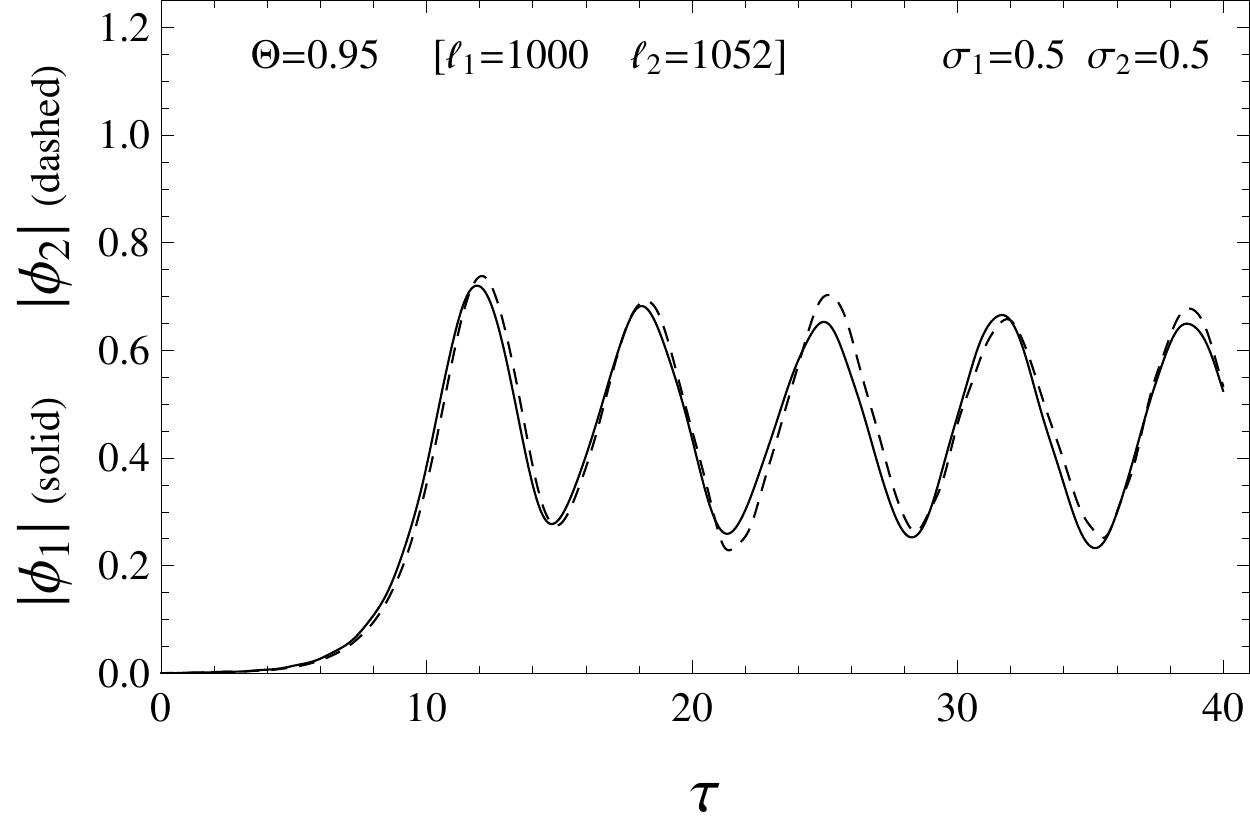}\qquad
\includegraphics[width=.35\textwidth,clip]{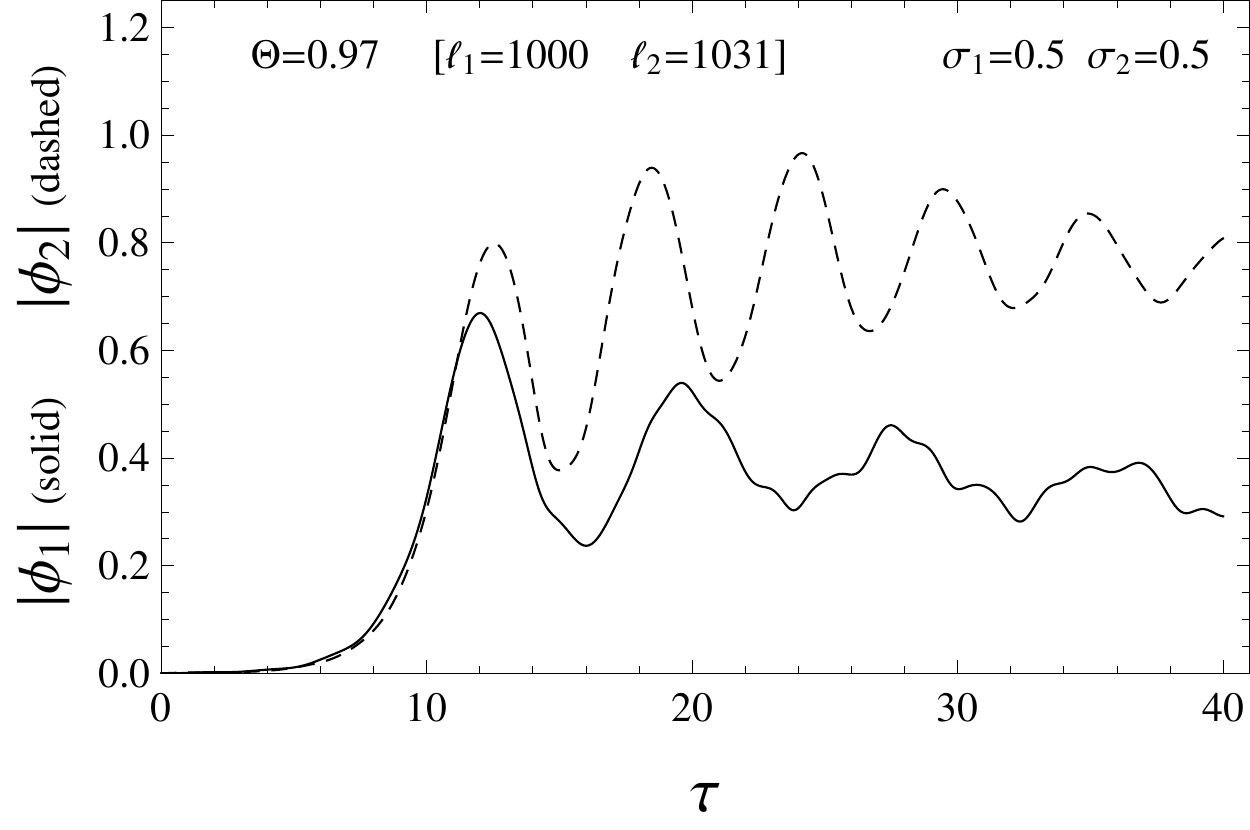}
\end{center}
\caption{Plots of the temporal evolution of $|\phija|$ determined by the system \reff{mainsys}. We have set two different values of the parameter $\params$ as indicated in the graphs and the simulations are run for $\su=\sd$. The initial conditions and the parameter values are discussed in the text.}
\label{fig:INTs}
\end{figure}

In what follows, these equations are solved via a Runge-Kutta (fourth order) algorithm assuming $N_1+N_2=10^5$ total charge particle sheets and $\etab=0.01$. For the considered time scales and for an integration step $h=0.01$, the total momentum and energy (derived from \erefs{totmomcons} and \reff{totenecons} written in the average velocity reference frame) are conserved with relative fluctuations of about $1\times10^{-3}$ and $4\times10^{-9}$, respectively. In \figref{fig:INTs}, we plot the resulting wave amplitude, obtained as a function of time, setting $\su=\sd=0.5$ in order give equal weight to the two beams. As initial conditions, we consider the fields of order $\mathcal{O}(10^{-2})$ and the sheets randomly distributed in $0<\bar{\xi}_{1,2}(0)<2\pi$, while the velocities at $\tau=0$ are $\bar{\xi}'_{1}(0)=-\bar{\xi}'_{2}(0)=\pi(\vu-\vd)/L\omp\etab=(1-\params)/2\lu\etab$. Finally, the values of $\beta_j$ from \eref{betabar} are now $\beta_j=\big[\lj(1+\params)/2\lu-1\big]/\etab$, with $\ell_2 = \ell_1/\Theta$.

From the behavior of the mode amplitudes, the presence of a threshold value of $\params$ clearly emerges, below which the two beams evolve independently generating two well separated single mode profiles. The transition to the interaction regime is anyway rather smooth.
\begin{figure}
\begin{center}
\includegraphics[width=.36\textwidth,clip]{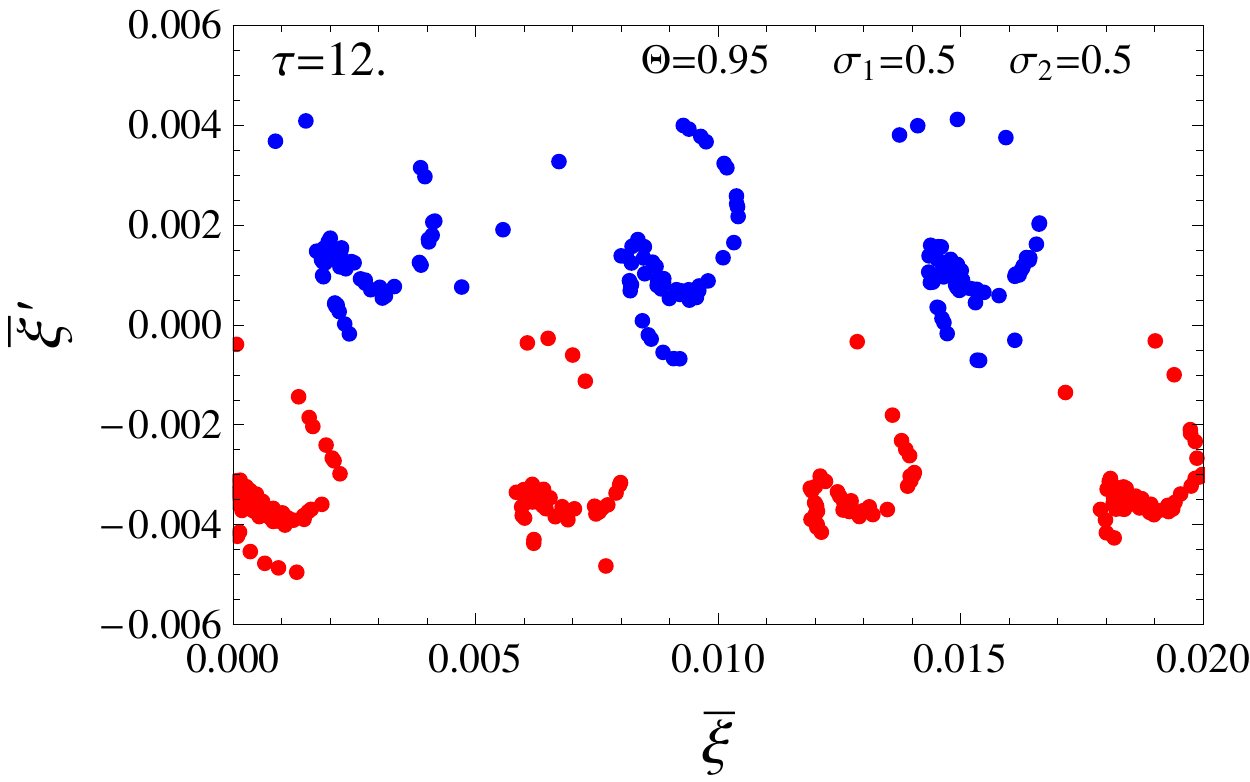}
\includegraphics[width=.36\textwidth,clip]{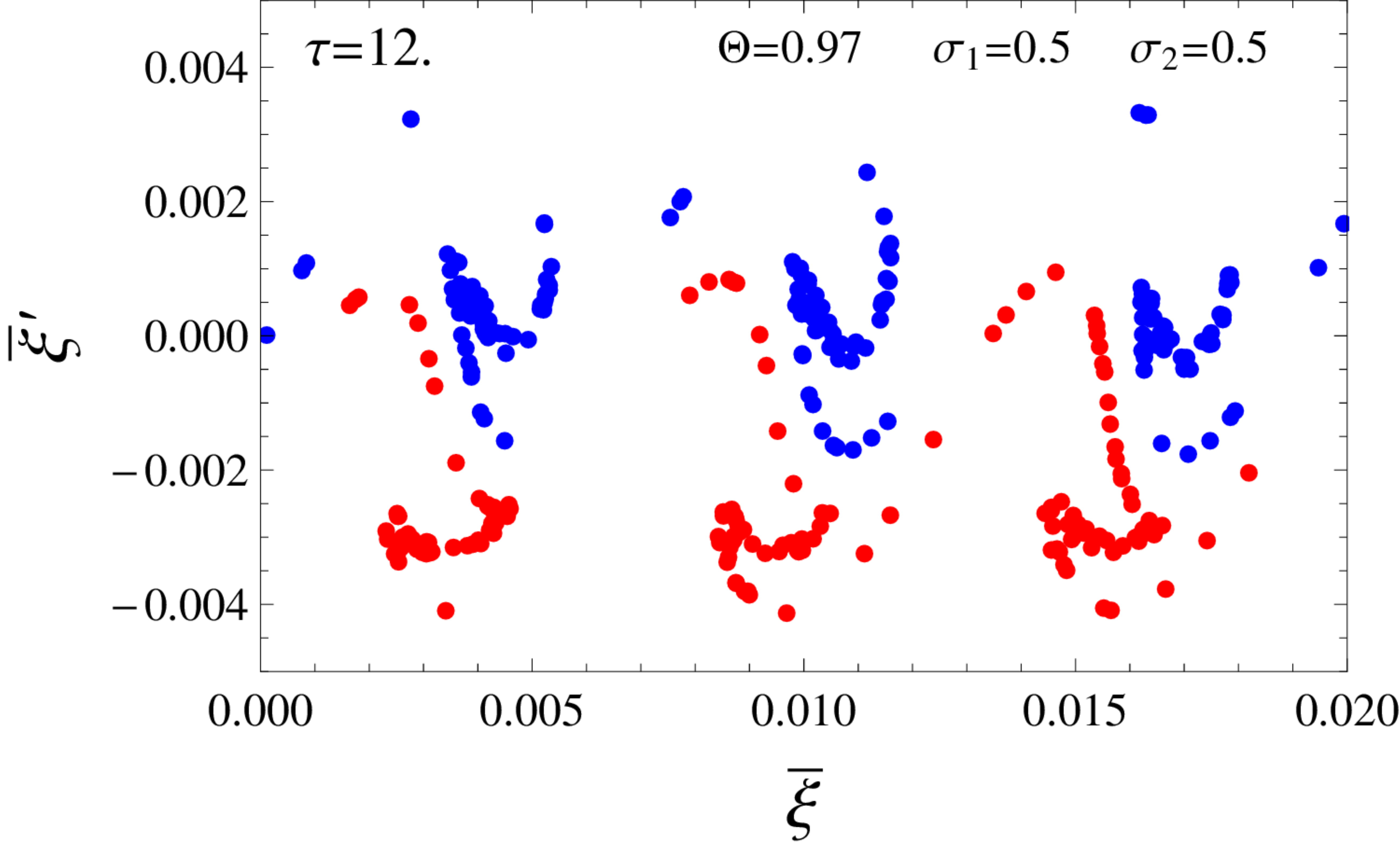}
\end{center}
\caption{Reduced phase-space $(\bar{\xi},\bar{\xi}')$ snapshots for fixed times (indicate in the graphs). Each point corresponds to a charge sheet labeled by $i$ in the system \reff{mainsys}, the beam $1$ ($2)$ is represented in blue (red), the beams have equal densities. We have set $\params=0.95$ (left-hand panel) and $\params=0.97$ (right-hand panel). (Color online)}
\label{fig:psloci}
\end{figure}

When $\params<0.96$, the generated field profile corresponds to two independent resonant modes. The amplification of both the waves proceeds as in the O'Neil analysis, where the initial exponential instability is followed by periodic oscillations. Furthermore, the instability is proportional to the corresponding $\sigma$ values consistent with linear theory. The dynamical independence of the beams can be argued also from the left-hand panel of \figref{fig:psloci}, where the phase space is depicted at a fixed time. Coherent with the single wave model, the dynamics of the system leads to the formation of rotating clumps which follow the corresponding Langmuir mode. In this sense, the number of such coherent structures is related to the wave number and we observe $\lu$ (beam $1$ in blue) and $\ld$ (beam $2$ in red) clumps in the periodicity length $L=2\pi$. The synchronized clumps evolve independently and no interference or transport between the two beams occurs. This behavior has been obtained in all the simulations for $\params<0.96$.

When $\params\gtrsim0.96$, the onset of interaction phenomena between the two beams is clear from \figref{fig:INTs}. Analyzing the phase-space dynamics in this case (in the right-hand panel of \figref{fig:psloci}, we set $\params=0.97$), we observe how during the formation of the rotating clumps, \ie during the initial exponential evolution of the mode, the charge sheets of the first beam are involved in the dynamics associated to the second one. This yields the evolution of the Langmuir waves described in \figref{fig:INTs} for the same parameter values, where the leading mode is indeed $\phi_2$. Such a mechanism is consistent with the fact that the mode $\phi_2$ is excited by the slower beam and can tap energy from the mode $\phi_1$ via wave-particle interactions with the faster beam particles.

\begin{figure}
\begin{center}
\includegraphics[width=.35\textwidth,clip]{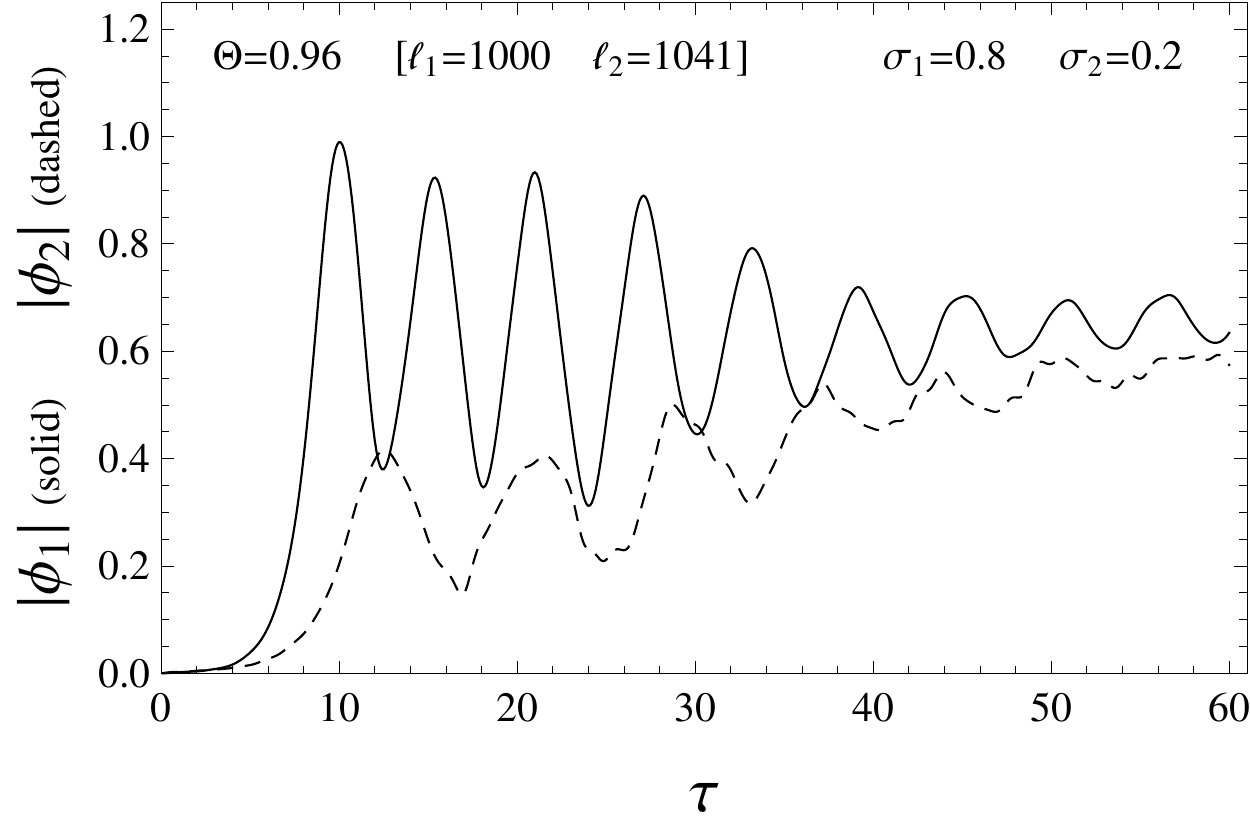}
\end{center}
\caption{Plots of the temporal evolution of $|\phija|$ by setting $\params=0.96$ and $\su=0.8$, $\sd=0.2$.}
\label{fig:surf}
\end{figure}

\begin{figure}
\begin{center}
\includegraphics[width=.3\textwidth,clip]{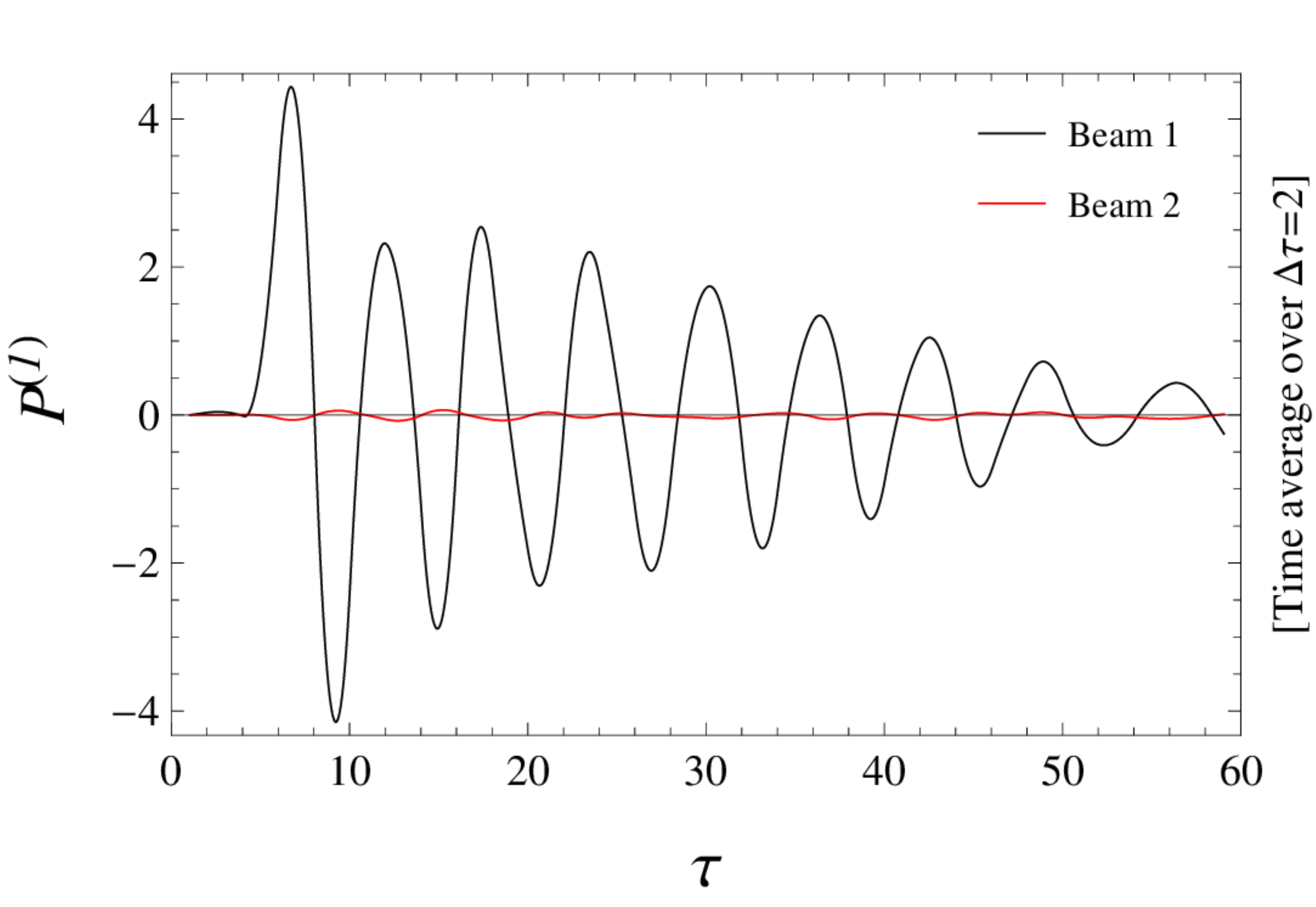}
\includegraphics[width=.3\textwidth,clip]{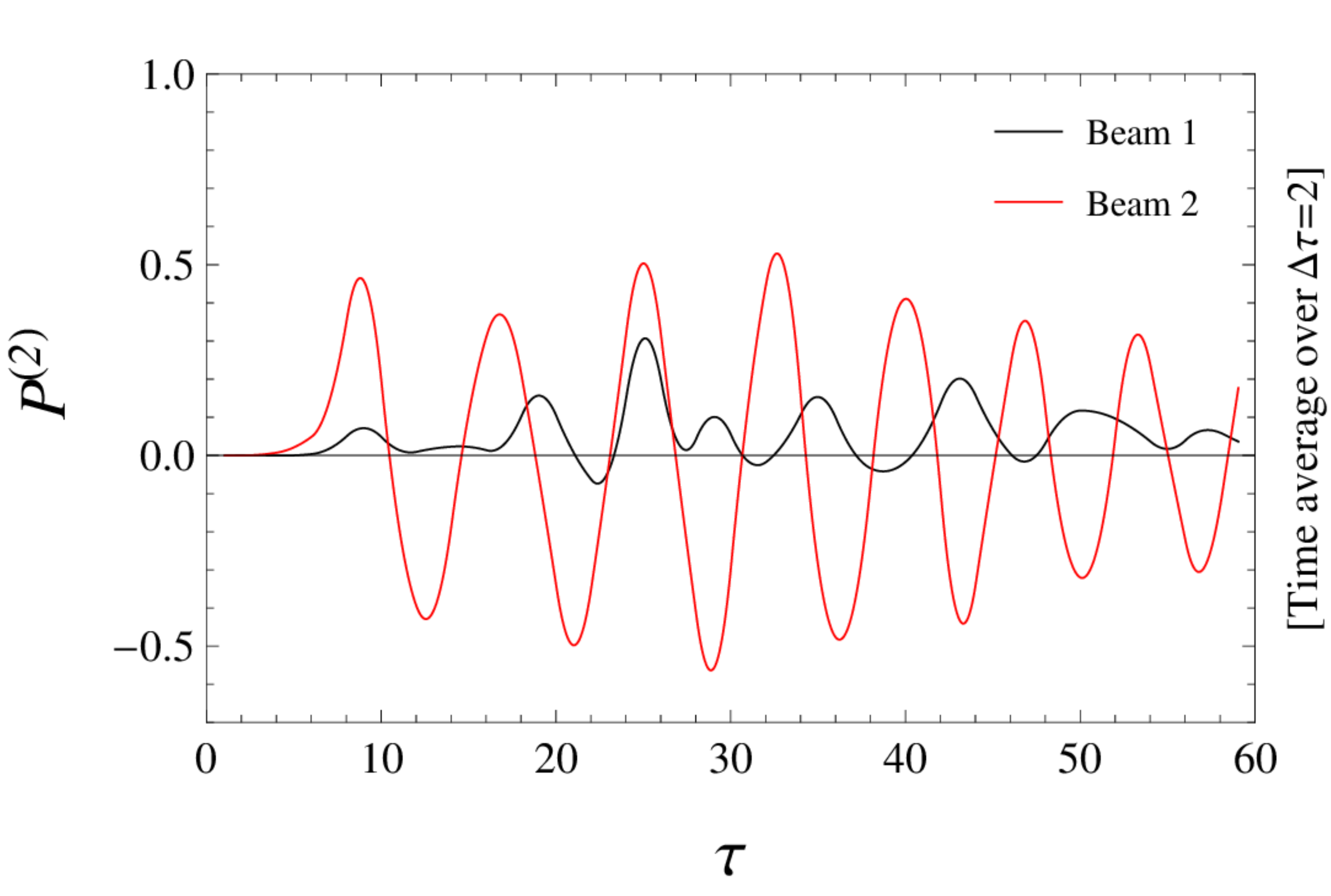}\\
\includegraphics[width=.3\textwidth,clip]{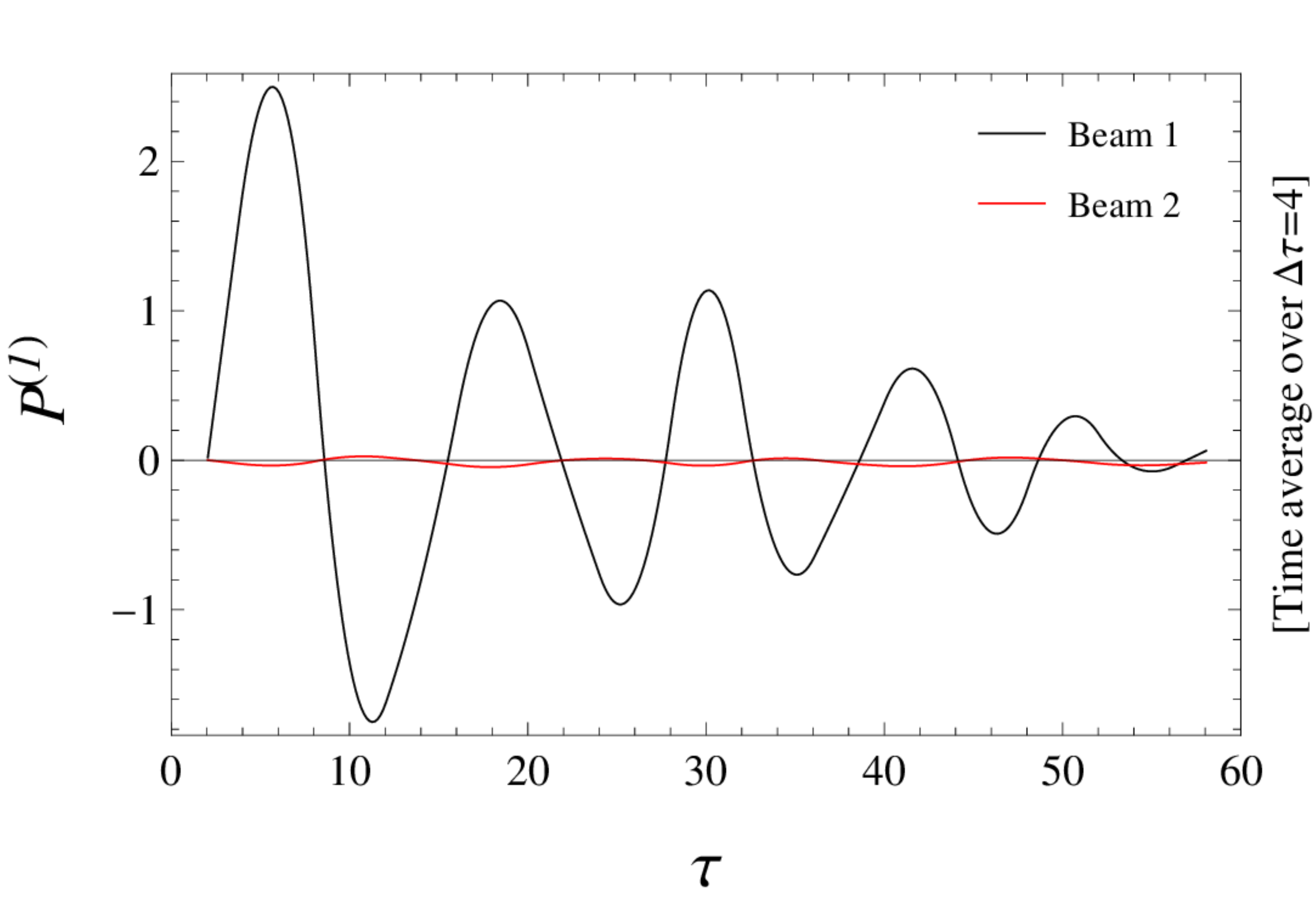}
\includegraphics[width=.3\textwidth,clip]{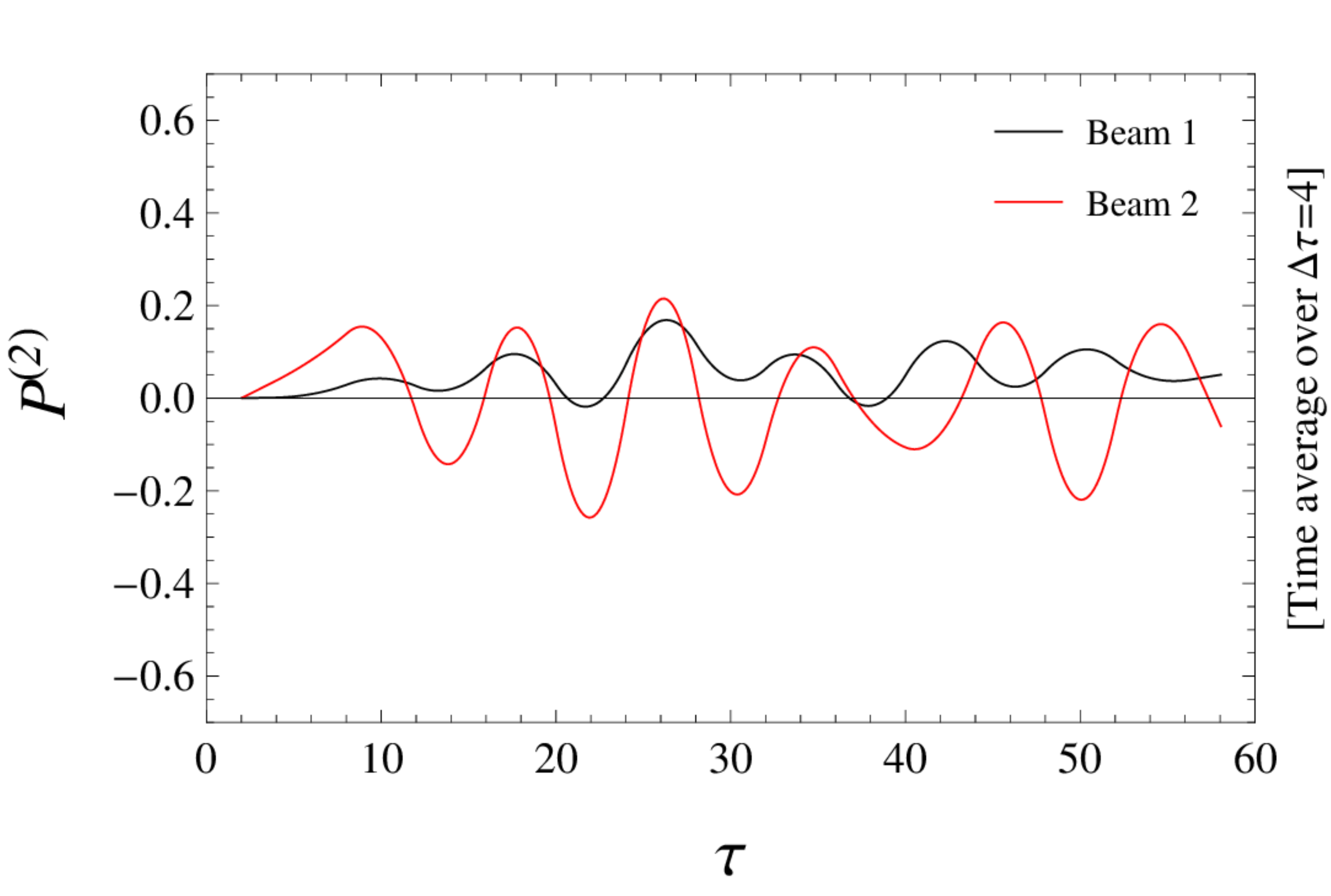}
\end{center}
\caption{Temporal evolution of $P_a^{(j)}$ (in the left-hand panels, the mode $j=1$; in the right-hand panels, the mode $j=2$) by setting $\params=0.96$ and $\su=0.8$, $\sd=0.2$. The power exchanged by the beam $a=1$ (2) is plotted in black (red). In the upper (lower) panels, we have time averaged the power over a step $\Delta\tau=2$ ($\Delta\tau=4$) in order to smooth out small rapid oscillations. The transferred energy to the mode $\ell_2$ from the two beams is comparable, indicating the extraction of power from the mode $j=1$ to the mode $j=2$. (Colors online)}
\label{fig:power}
\end{figure}

Interaction phenomena are evident also in the case of different values of the beam densities. In this respect, it is worth noting that a peculiar feature emerges near the threshold $\params$-value when $\su$ is significantly greater than $\sd$, see \figref{fig:surf}. The larger density of the first beam forces the second mode to evolve differently compared to the equal density case. In particular, the weaker mode $\phi_2$ is increased by the presence of the other one, again, because of energy transfer mediated by wave-particle interactions. This mechanism is analogous but opposite to the ``surf'' phenomenon occurring in plasma acceleration processes \citep{Li14,ES96,KJ86}. Specifically, when the second beam velocity $\vd$ is close to the enhanced Langmuir mode phase velocity, $\phi_2$ continues to tap energy from the faster beam. Thus, the weaker/slower mode can grow at the expense of the stronger/faster mode, via power transfer mediated by wave-particle interactions.
In order to clarify the mechanism underlying this phenomenon, we can plot the power exchanged between each beam and individual modes, in the case under consideration ($\params=0.96$ and $\su=0.8$, $\sd=0.2$). For the beam $a$ and the mode $j$ the dimensionless power can be written, using the scaled variables in the reference frame of the average beam speed, as:
\begin{align}
P_a^{(j)}=
\sum_{i=1}^{N_a}\Big(
i\,\ell_j^{-1}\,{\phi}_j\;\;e^{i\ell_j\bar{\xi}_{ai}+i\beta_j\tau}
\Big(\bar{\xi}_{ai}'+\frac{\pi(\vu+\vd)}{\omp\etab L}\Big)+c.c.\Big)\;.
\end{align}
This quantity is plotted in \figref{fig:power} as a function of $\tau$ separately for the two modes and with two different time averages as indicated in the figure, in order to better illustrate the global power transfer features. It is easy to realize from the right-hand panels that, when the ``anti-surf'' mechanism begins ($\tau\sim20$, see \figref{fig:surf}) the power exchanged by the two beams is comparable. Indeed, the particles of the fastest beam transfer energy to the mode $\ell_2$ and drain energy from the mode $\ell_1$. This behavior of power transfer confirms and clarifies the time evolution of the mode amplitude sketched in \figref{fig:surf}.

\begin{figure}
\begin{center}
\includegraphics[width=.35\textwidth,clip]{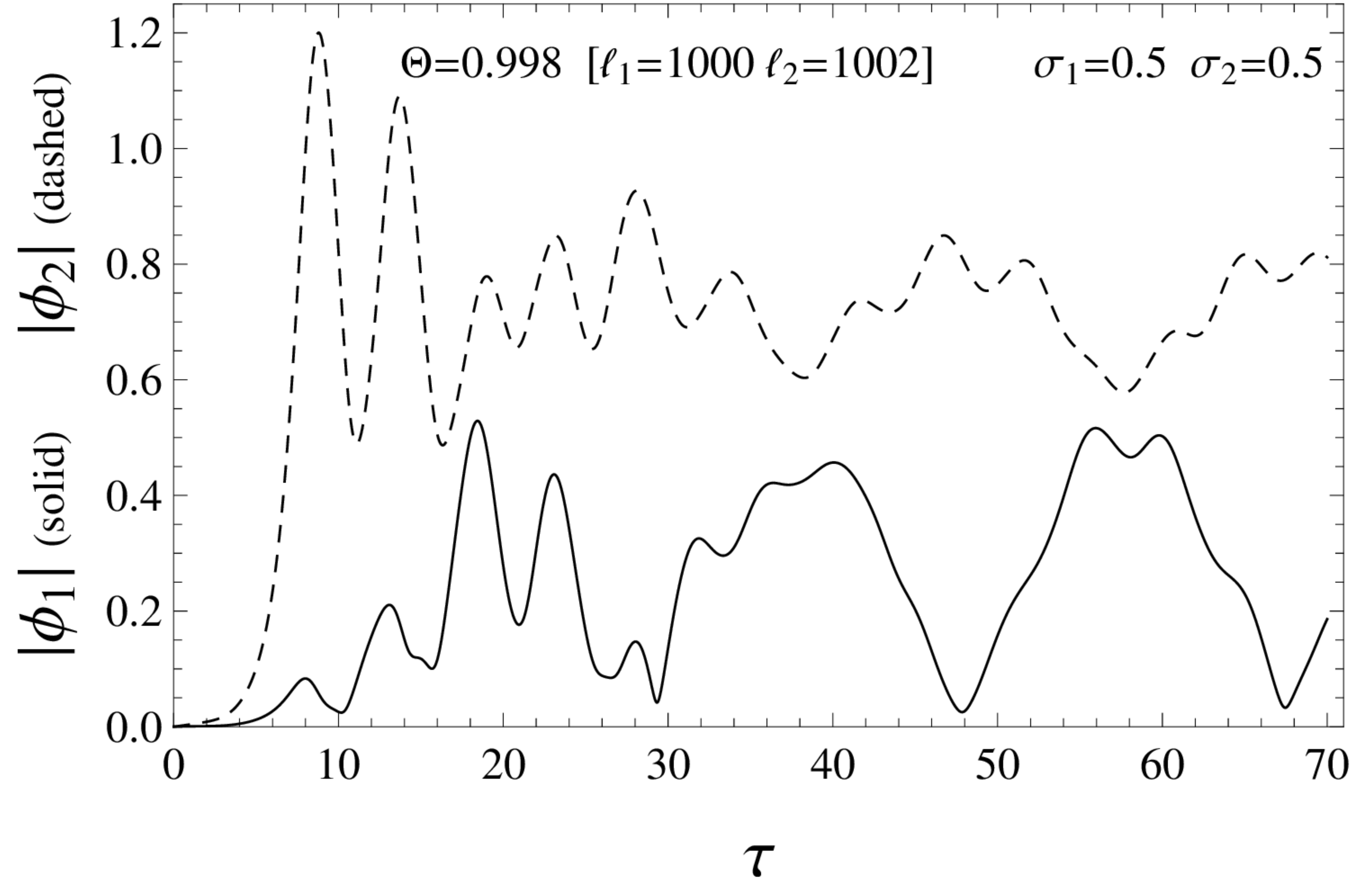}
\end{center}
\caption{Plots of the temporal evolution of $|\phija|$ by setting $\params=0.998$ and $\su=0.5$, $\sd=0.5$.}
\label{fig:huisghgskjh}
\end{figure}

We conclude by briefly discussing the limit case for $\params$ of order $\etab$. In particular, when $\params\gtrsim0.995$ we observe the onset of a very particular behavior of the mode nonlinear evolution, as indicated in \figref{fig:huisghgskjh} where we have set, as an example, $\Theta=0.998$. This is due to the fact that the beam instabilities overlap and the resulting dynamics is completely different from the previous cases. As soon as the clumps form, the beam charge-sheet interference is strong and the two beams can be effectively treated as a single phase-space structure but in the presence of peculiar time-dependent potentials generated by a fully interacting regime.

\subsection{Quasi-degenerate modes}\label{quasidegen}
Let us now analyze how, in addition to the two most unstable waves discussed above, the two beam system can excite other nearly degenerate modes, \ie fluctuations at frequency $\omega_j\simeq\omp$ implying $\eps\simeq0$. 

For $\params$ over the threshold value for efficient nonlinear beam coupling, these modes correspond to sidebands, defined as $k^{+}_j=k_2+p\Delta k$ and $k^{-}_j=k_1-p\Delta k$ (where, $\Delta k=k_2-k_1$ and $p$ is an integer number). These quasi-degenerate modes are responsible for spectral broadening and, possibly, for a cascade phenomenon. To illustrate this process, we now discuss the case $p=1$, considering thus the two sideband modes $k^{+}_3=2\kd-\ku$ and $k^{-}_3=2\ku-\kd$. In the reference frame of the average beam speed, the dynamical equations governing the system can be easily constructed generalizing \erefs{mainsys} for the two additional quasi-degenerate modes. The corresponding mismatch parameters are $\beta^{+}_3=\big[1/\params-\params/2-1/2\big]/\etab$, $\beta^{-}_3=\big[\params-1/2\params-1/2\big]/\etab.$
\begin{figure}
\begin{center}
\includegraphics[width=.23\textwidth,clip]{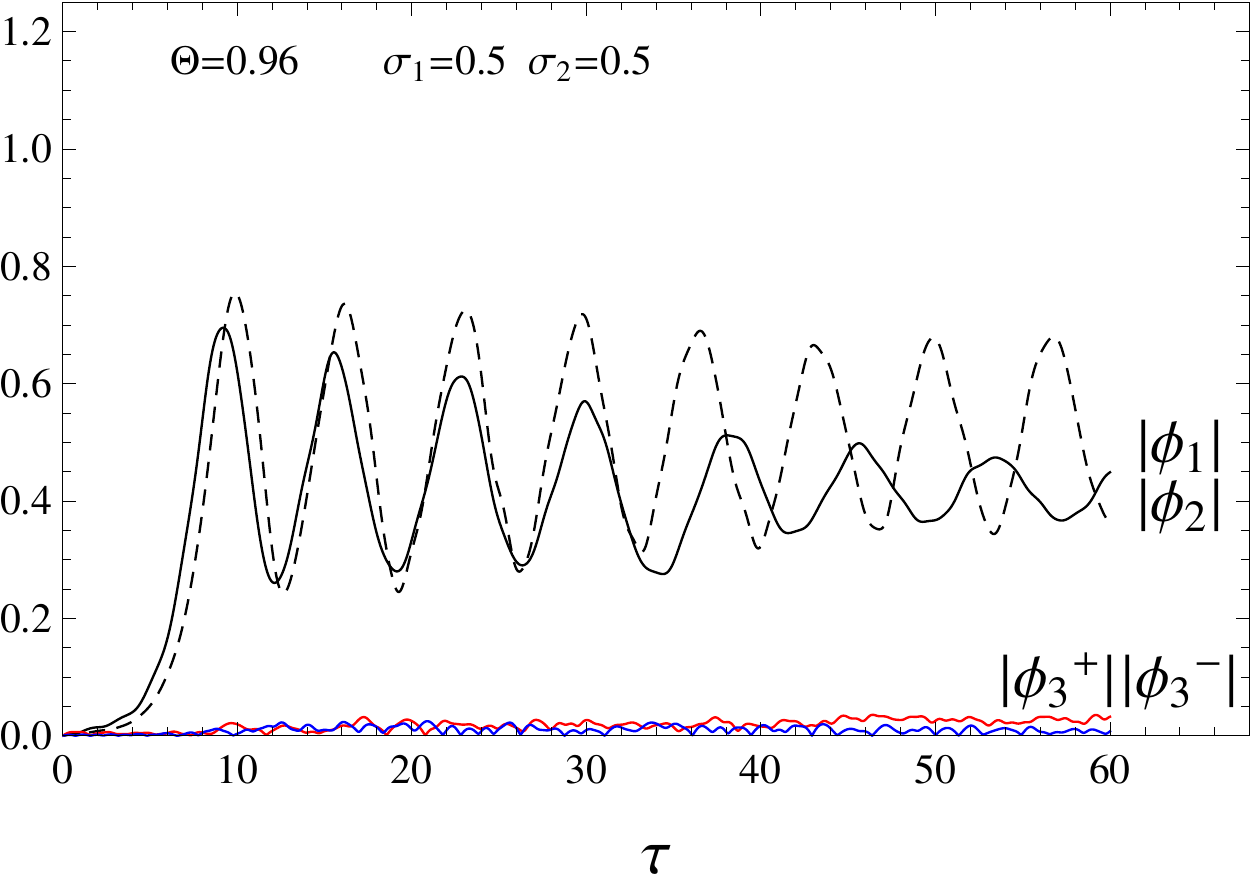}
\includegraphics[width=.23\textwidth,clip]{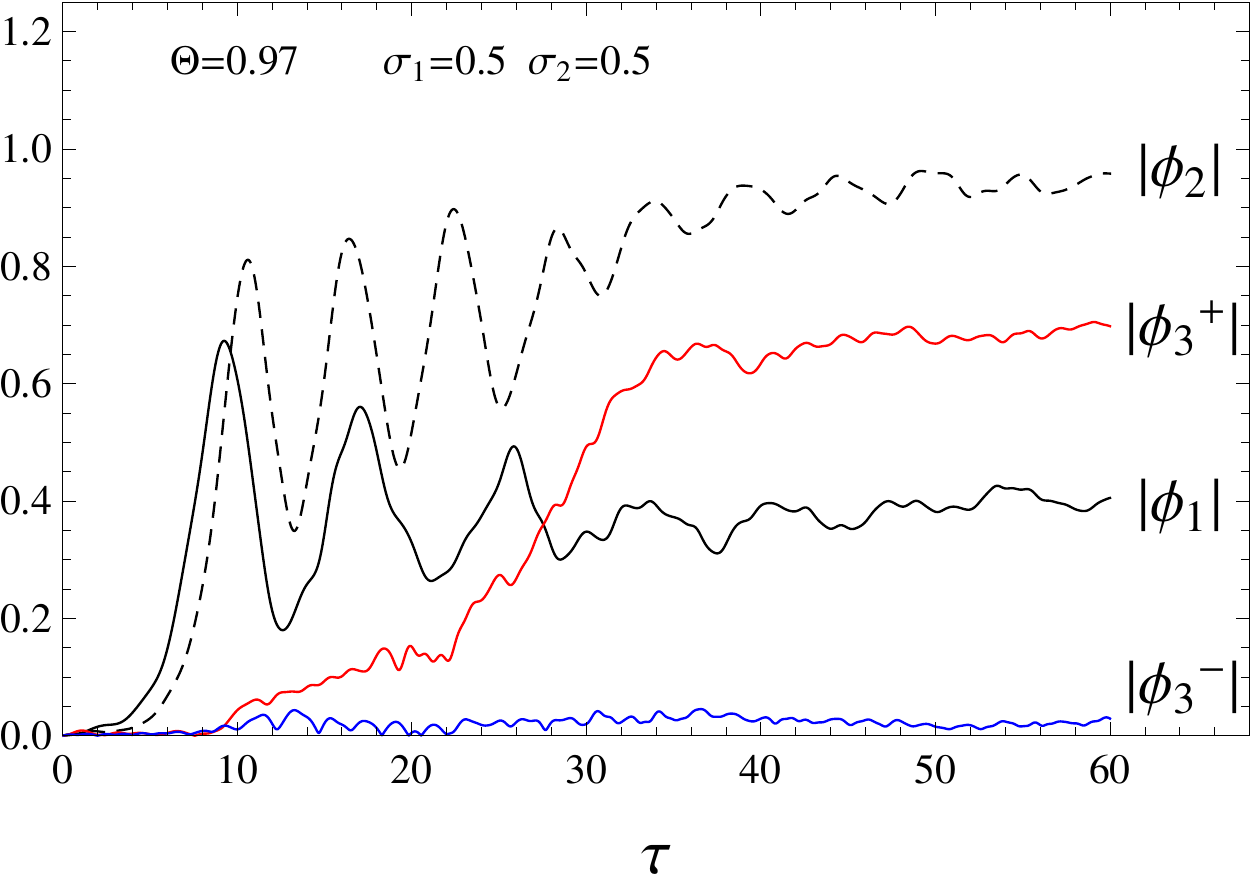}
\includegraphics[width=.23\textwidth,clip]{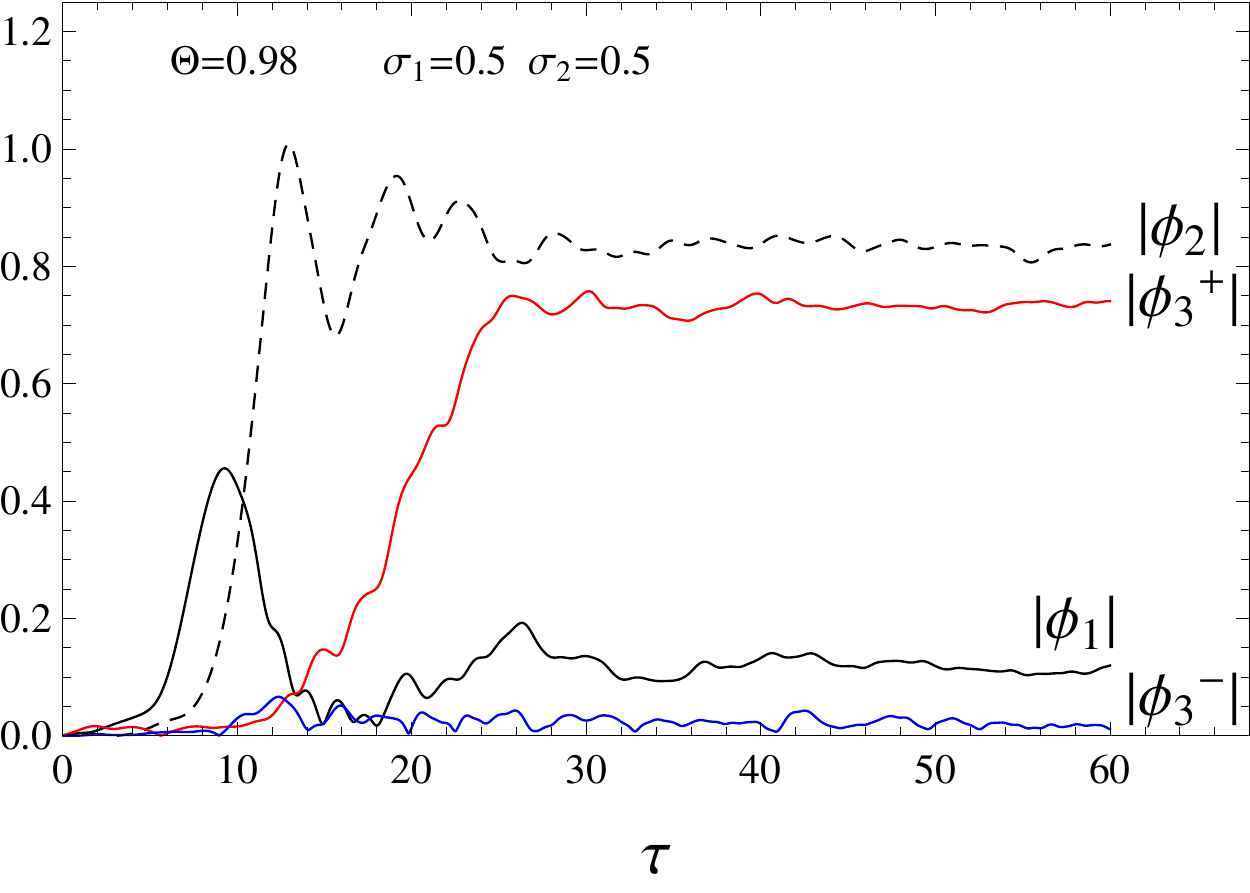}
\includegraphics[width=.23\textwidth,clip]{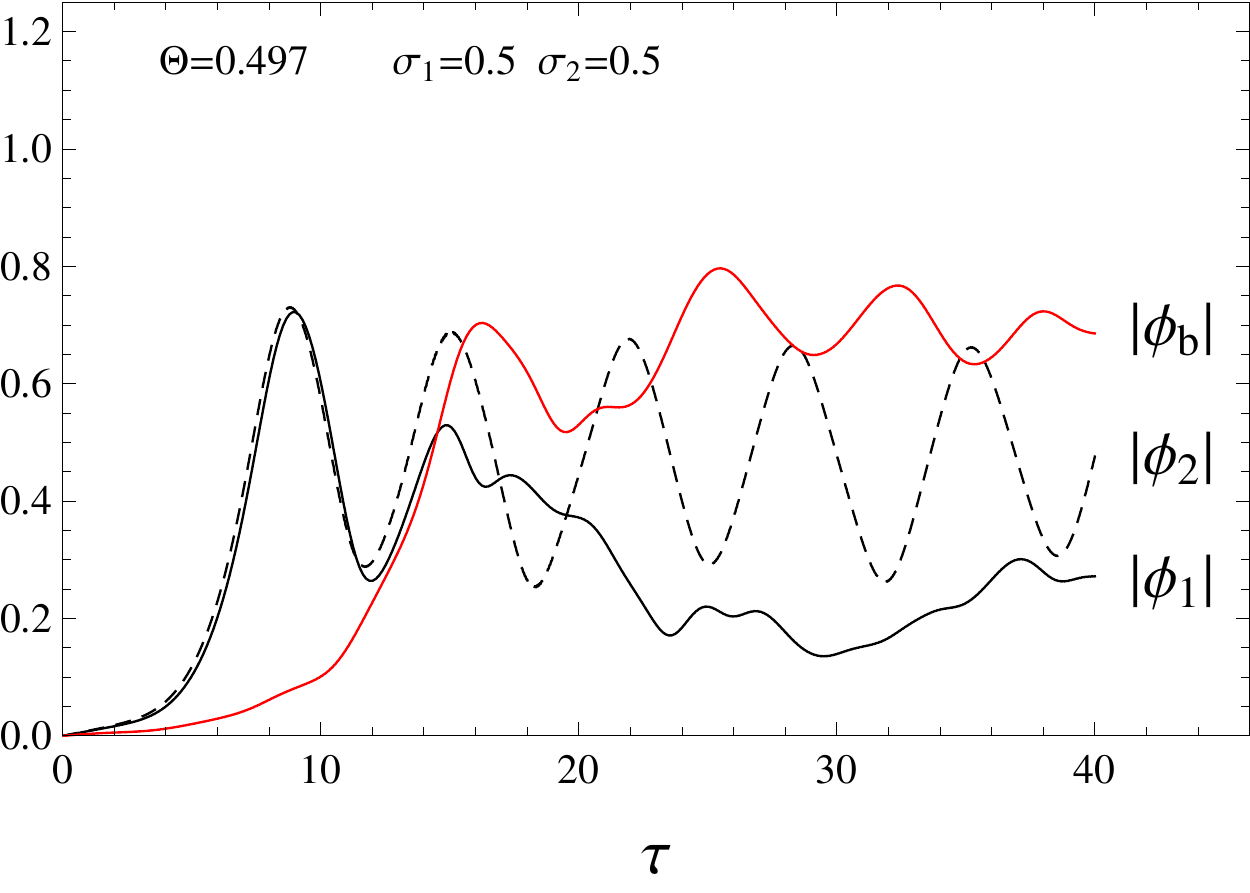}
\end{center}
\caption{First three panels from the left: Plots of the temporal evolution of $|\phi_1|$, $|\phi_2|$, $|\phi^{+}_3|$ and $|\phi^{-}_3|$ for different values of $\params$ as indicated in the graphs. Right panel: Plot of the mode $\phi_b$ as function of time. For this figure, we set $\su=\sd$ and $\etab=0.01$. (Color online)}
\label{fig:sideband1}
\end{figure}

From the first three panels of \figref{fig:sideband1}, only $k^{+}_3$ mode can be excited, while $k^{-}_3$ does not grow since its resonance condition corresponds to velocities larger than $v_1$ and, further, it is associated with a negative value of the dielectric function (its frequency is actually slightly smaller than $\omp$). It is important to note that the unstable mode at $k^{+}_3$ saturates at comparable amplitude to the resonant mode $\phi_2$ although saturation is reached at a later time.
Meanwhile, $k_3^+$ sideband excitation can occur only above the threshold condition in $\Theta$ for efficient nonlinear beam coupling. In fact, the mechanism for sideband excitation is the same as that for power transfer from faster to slower resonant modes discussed above.

Efficient sideband generation can also occur in the case of two independent resonant beam modes, under special limiting conditions. This is the case when $\Theta \rightarrow 1/2^-$; \ie $k_2 \rightarrow 2k_1^+$ and the beat mode $k_b \equiv k_2-k_1 \rightarrow k_1^+$. In fact, in this case, illustrated in the right-hand panel of \figref{fig:sideband1}, the $k_b$ mode ($\phi_b$ in the plot) is quasi-degenerate with the Langmuir frequency $\omp$ and has a resonance condition at velocities slightly below $v_1$. Thus, it must be treated nonlinearly on the same footing of the two most unstable modes. As shown in the figure, the saturation level of the quasi-degenerate mode approaches the same value of $\phi_2$ saturation, while the evolution of $\phi_1$ is significantly altered because $\phi_b$ absorbs a relevant part of its energy via nonlinear wave-particle interaction. It can be also demonstrated that $\phi_b$ is not efficiently exited as $\params\to1$. This fact is consistent with the original problem studied by \citep{OWM71}, where mode-mode interactions are negligible.

\section{Lagrangian Coherent Structures and the transport processes}\label{ftlesection}
In this Section, we analyze the transport processes in order to identify the regions where mixing phenomena are fastest and, in particular, their boundaries that are known as LCSs. The LCS technique was introduced by G. Haller in order to generalize intrinsically asymptotic concepts in the dynamical system theory, such as stable and unstable manifolds, to the study of systems over a finite time span $\Delta\tau$. In the same way the separatrix of a pendulum splits the phase space into two macro regions where initial conditions have different asymptotic behavior, \ie rotations and librations, the LCSs of the beam-plasma system split the particles into two groups, \ie recirculating and non-recirculating ones over the finite time span $\Delta\tau$. Without entering into the mathematical details of the definition, following Haller \citep{Haller15l} we state that LCSs need to be defined as material lines, \ie lines advected by the flow, which are locally most repulsive with respect to the nearby ones over the time span $\Delta\tau$. Two initial conditions taken on different sides of an LCS will increase their distance over the time span $\Delta\tau$ more than two initial conditions on two sides of any other curve. Therefore they will have, in general, a qualitatively different evolution the same way that two initial conditions starting on different sides of the stable manifold of a pendulum will increase their distance more than two initial conditions taken on the same side. This leads to the definition of a functional, called repulsion rate, which needs to be maximized on the LCS trough a variational approach \citep{H11,MPJPP}. This technique has already been used in the computational fluid dynamics community \citep{H01,SLM05,Se11,Sa12,Ga07} and, more recently, in plasma physics \citep{Bo11a,Bo11b}. An introductory discussion on the proper operational definition of these structures can be found in \citep{MPJPP}. Meanwhile, for a different approach to the characterization of coherent structures, see \citep{LZ02}.

\subsection{General approach}
In the recent review \citep{Haller15l}, the author states that an LCS can be sought as an extremizing curve of the FTLE field, \ie a ridge. Actually, a debate exists in the  literature about the proper definition of a ridge. We refer the interested reader to \citep{Haller15l} or to \citep{MPJPP} for more in-depth discussions of applications of this method in plasma physics. In our analysis, following \citep{Haller15l}, we define a ridge of the FTLE field as a curve along which one sees a smaller change in the value of the field than in the direction transverse to it. The resultant contour plot of the FTLE is very peaked and almost constant on nearly 1D regions. We will identify these structures with the ridges of the FTLE field and, therefore, with the LCSs of the system \citep{LR10,KPHHNT10,VHG02,CRASSWY14,MHPRS07}.

In order to characterize the FTLE fields for a given configuration, the trajectories of several test particles are traced when evolved in the time-dependent potentials $\bar{\phi}(\bar{\xi},\tau)$ (using the notation of the two-beam problem) generated self-consistently by the charge distribution. For a single reference time, the test particles are initialized to sample the whole phase space of interest (depending on the parameters). The system is composed of an arbitrarily large number of particles and, therefore, we can neglect the contribution of a single test-particle in the modification of the potentials, which are treated as assigned for a correspondent beam-plasma simulation. This approximation has already been used in \citep{SP78,TMM94}.

To evaluate the FTLE profile for a fixed time $\tau$, the test particles are initialized in two phase-space grids of $M/2$ particles each and having an initial infinitesimal displacement in the velocity direction. In other words, a test particle located in $[\bar{\xi},\bar{\xi}']$ has a neighbor initialized in $[\bar{\xi},\bar{\xi}'+\delta_\tau]$. With the assigned time-dependent potentials, at a time $\tau+\deltat$ the considered two test particles will be at a distance $\delta_{\deltat}$ in the phase-space and the FTLE value $\sigma$ in the point $[\bar{\xi},\bar{\xi}']$ is evaluated as \citep{H11}
\begin{equation}\label{eqftle}
\sigma(\bar{\xi},\bar{\xi}',\tau,\deltat)=
\ln\,(\delta_{\deltat}/\delta_\tau)/\deltat\;.
\end{equation}
When considering a positive time evolution $\deltat>0$, the curves where the FTLE field is peaked define a repulsive transport barrier, while setting $\deltat<0$ and evolving the system backward in time they represent an attractive barrier. Approximated 1D structures which correspond to the LCSs can be built by plotting the maximum values of $\sigma(\bar{\xi},\bar{\xi}',\tau,\deltat)$ as extracted from a contour plot. In the analysis of this Section, we have set two grids of $250\times250$ test particles ($M=125000$), thus obtaining $62500$ values of the FTLE for each phase-space snapshot. The self-consistent fields $\bar{\phi}(\tau)$ have been sampled from the complete Hamiltonian simulation with the proper selected parameters.

\subsection{The single wave model}\label{sec:singlewavemodel}
As first step, we apply the FTLE method to characterize the dynamical system describing the single wave model \citep{OWM71}, which can be easily deduced from \erefs{generalsystem} by considering one beam and one mode $j=1$. In \citep{TMM94} and \citep{SP78}, the test-particle approximation is used for studying the behavior of the single wave model using the dynamical systems techniques. In particular, in the latter, the evolution of the electromagnetic field is modeled with an analytic function while, in the former, it is extracted from the numerical simulations as in the present paper. In \citep{TMM94}, the periodicity of the asymptotic state of the system is exploited in order to generate a Poincar\'e section (stroboscopic map) of the motion, \ie the sampling of the evolution of a particle after each periodic oscillation of the electromagnetic field. The set of points obtained by this procedure, for different initial conditions, can be used to distinguish different kinds of motion: periodic, quasi-periodic or chaotic. In both works mentioned above, a region of non-chaotic motion is present. This region moves, with a periodic behavior, coherently, \ie only with slight deformations in its shape. Some of the beam particles will start inside this structure and, thus, they will be trapped during their entire evolution. This is the mechanism responsible for the formation of the clump in the Hamiltonian system and for the bunching of the particles. This procedure splits the ensemble of trajectories and, thus, the ensemble of the available particle motions, relying on their asymptotic behavior. For this reason, this scheme identifies barriers for the transport processes in the system which holds at any time, \ie curves which cannot be crossed by the particles during their motion. In the most general aperiodic case however, such structures do not exist because every trajectory could in principle be chaotic and, furthermore, it is not clear how the stroboscopic map technique can be used for studying the single wave model during the linear growth.
\begin{figure}
\begin{center}
\includegraphics[width=.28\textwidth,clip]{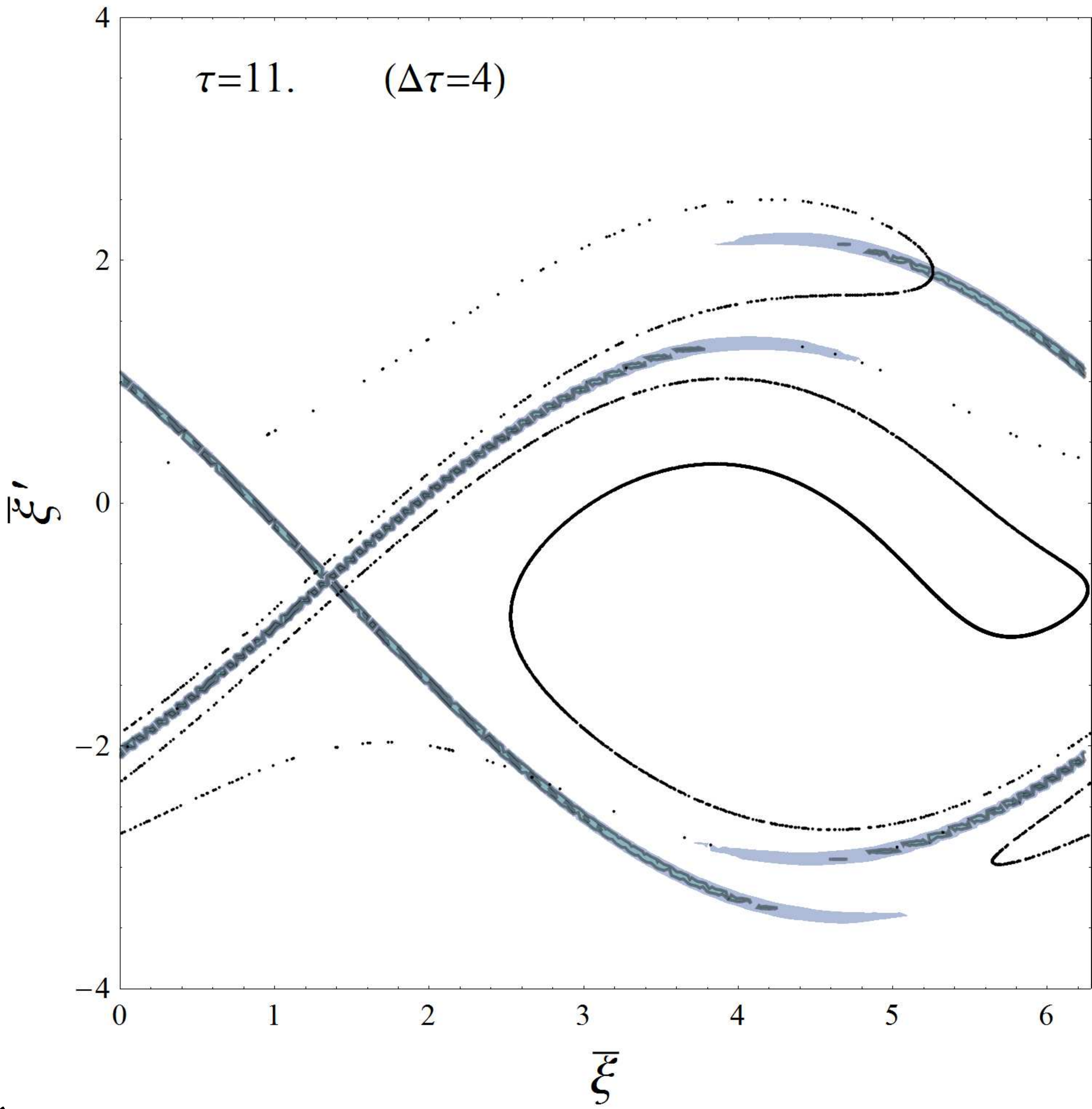}
\includegraphics[width=.28\textwidth,clip]{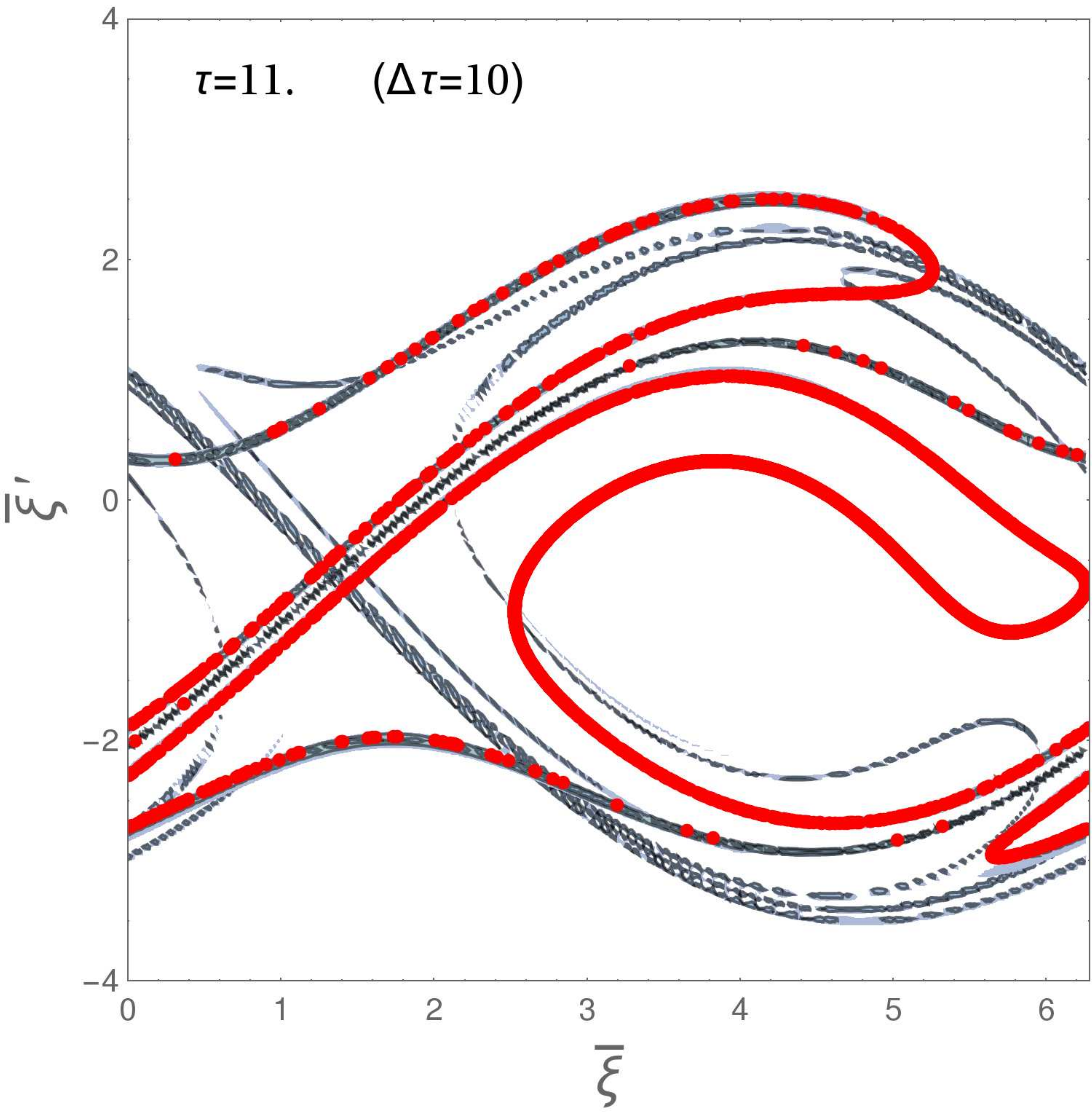}
\includegraphics[width=.28\textwidth,clip]{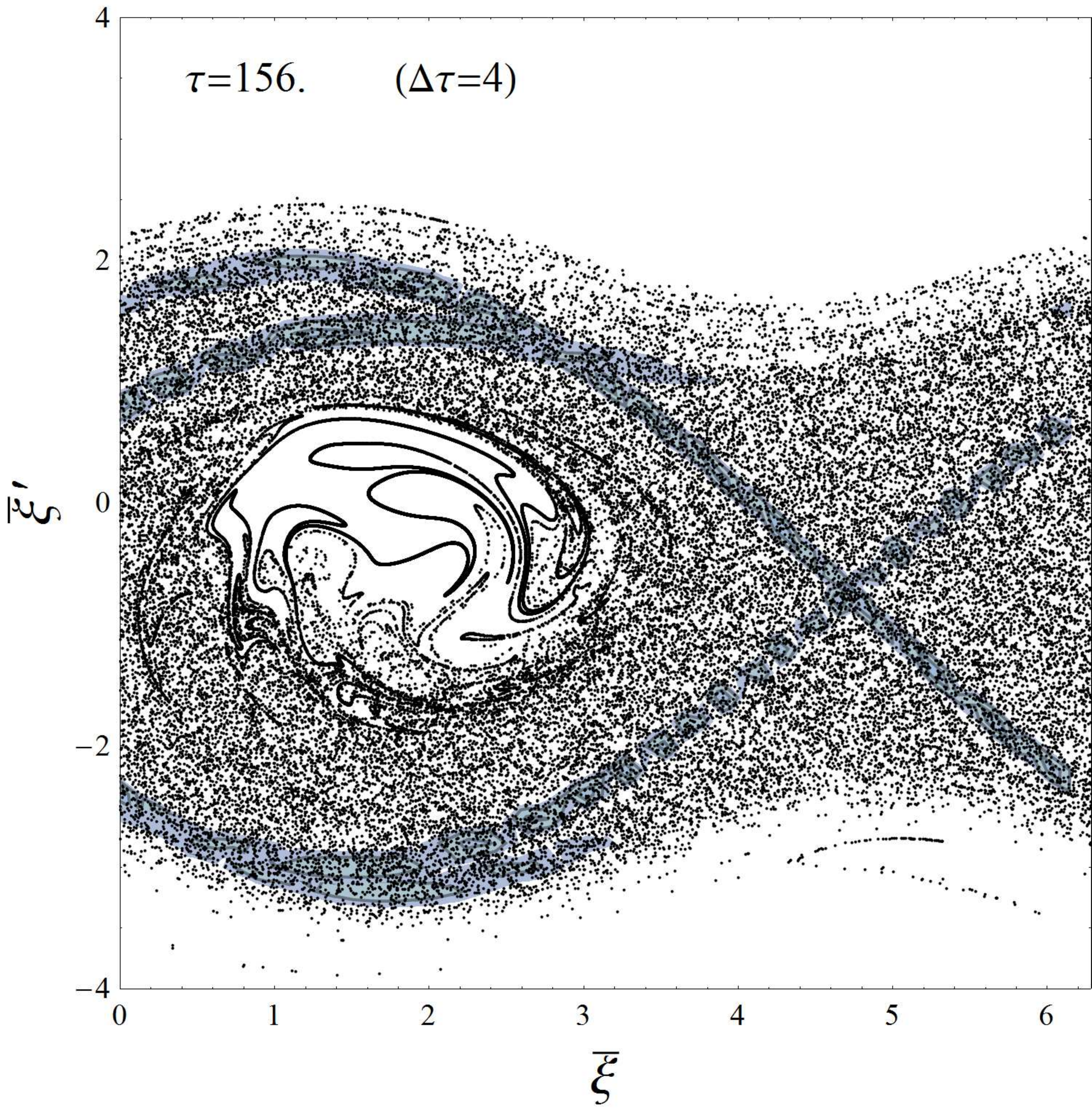}
\end{center}
\caption{Overlay of the phase-space snapshots (black or red points) and the LCSs (thick lines) determined as the peaks of the FTLE fields extracted from a corresponding contour plot built using \eref{eqftle}. Left panel: LCSs for the single wave model for $\tau=12.5$ and $\deltat=4$. Middle panel: LCSs for $\tau=11$ and $\deltat=10$ (in this plot, the beam particles are highlighted in red). Right panel: LCSs in the late evolution for $\tau=156$ and $\deltat=4$. (Color online)}
\label{fig:ftleswm}
\end{figure}

In this work, we are interested in studying the system during the onset of the instability. For this reason, the periodicity of the electromagnetic potential is generally lost. Using the FTLE method in order to identify the shapes of the LCS can nonetheless split the phase space into different regions based on the behavior of the trajectories over a finite period of time. These structures, which are characterized by a peaked FTLE field, approximate the unstable (stable) manifolds of the dynamical system and characterize the transport through the lobes dynamics \citep{MW98}.

The two parameters that characterize the FTLE values are $\tau$ and $\deltat$. The time evolution of the LCS is identified by the first, while the second determines how refined the LCS will be. By choosing an increasingly longer integration time $\deltat$, the obtained LCSs acquire more intricate structures because of an increasingly better approximation to the homoclinic tangle. This kind of dynamics have been studied by \citep{MW98} and it leads to the splitting of the phase space into macro-regions which exchange a relatively small amount of particles over a characteristic time scale. The characterization of the detailed shape of the stable and unstable manifolds, which can be achieved trough an FTLE plot with long $\deltat$, gives insights into the time asymptotic transport processes. The shape of the LCS obtained with smaller values of $\deltat$, meanwhile, highlights the boundaries of macro-regions with small transport among them.

Here, we are interested in the shape and the location of these regions and, thus, the integration time is fixed as $\Delta \tau = 4$ after a number of tests, obtaining a simple, closed region as shown in the left-hand panel of \figref{fig:ftleswm}, where we have over-plotted the beam-particle positions to the LCS. The effect discussed above of increasing $\deltat$ can be deduced by comparing with the middle panel. Increasing the value of $\tau$, we get the position of the LCS at different instants and, therefore, the motion of the ``finite time'' transport barriers. From these plots, it can be argued that, during the onset of the instability, most of the beam particles are trapped inside the central region delimited by the LCS and, therefore, they recirculate and start forming the clump.

In order to make a comparison with \citep{TMM94}, we can calculate the LCS for larger values of $\tau$, when the electromagnetic potential oscillates periodically. We obtain the plot on the right-hand panel of \figref{fig:ftleswm}, where the transport barriers are splitting the chaotic region into two different areas having qualitatively different motions on a short time scale. The points outside the central ``eye'' have a motion similar to rotations in a pendulum, while the points inside the ``eye'' resemble librations. In contrast to an unperturbed pendulum, the points may start undergoing rotational motion and later undergo librational motion and vice-versa. However, this transition has a typical ``long'' time scale and, if we look at the transport processes which happen over a shorter time, the two different kinds of trajectories do not mix. For this reason the LCS may be regarded as finite time transport barriers.

\subsection{The two-wave model}
\begin{figure}
\begin{center}
\includegraphics[width=.23\textwidth,clip]{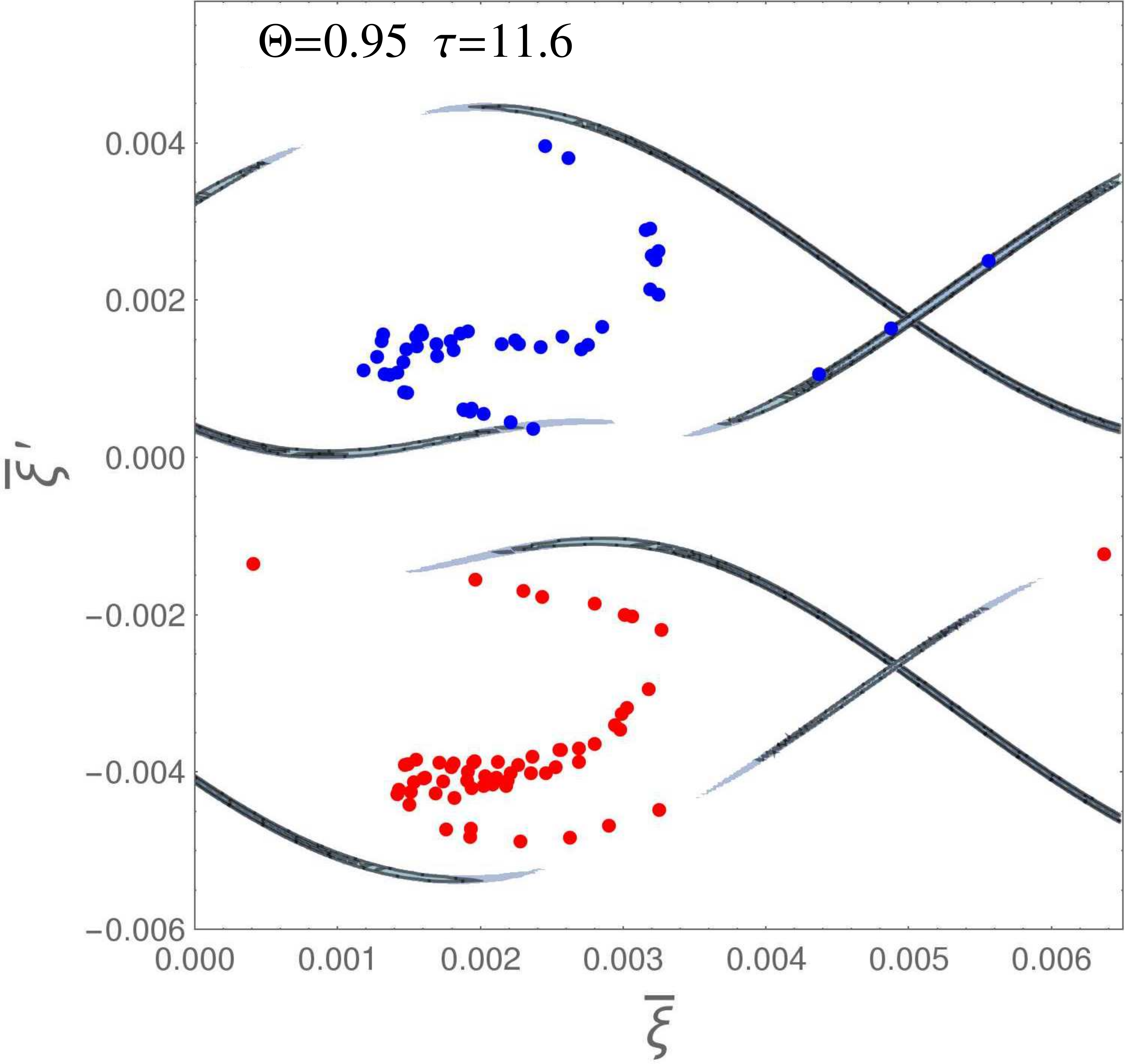}
\includegraphics[width=.23\textwidth,clip]{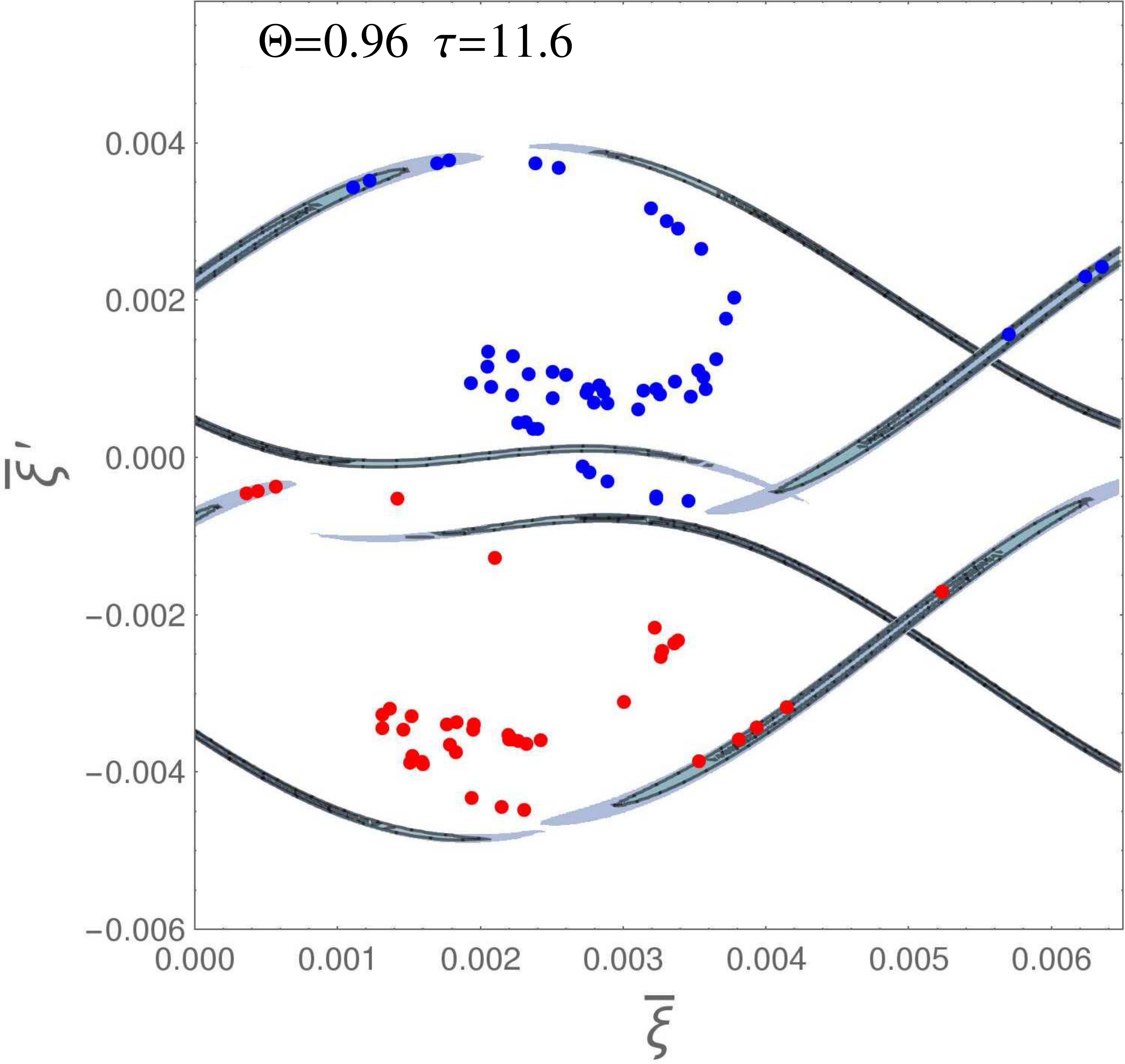}
\includegraphics[width=.23\textwidth,clip]{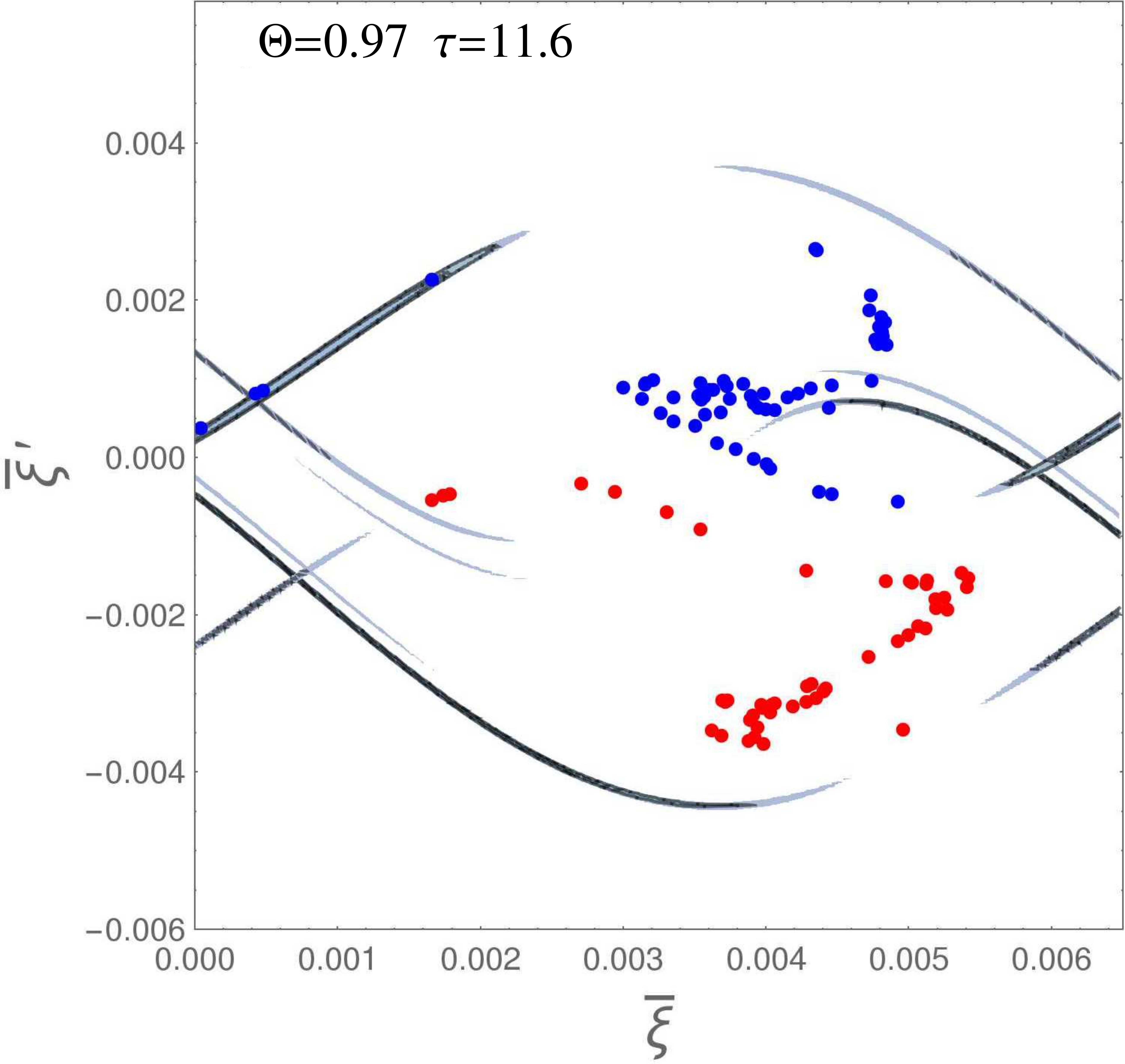}
\includegraphics[width=.23\textwidth,clip]{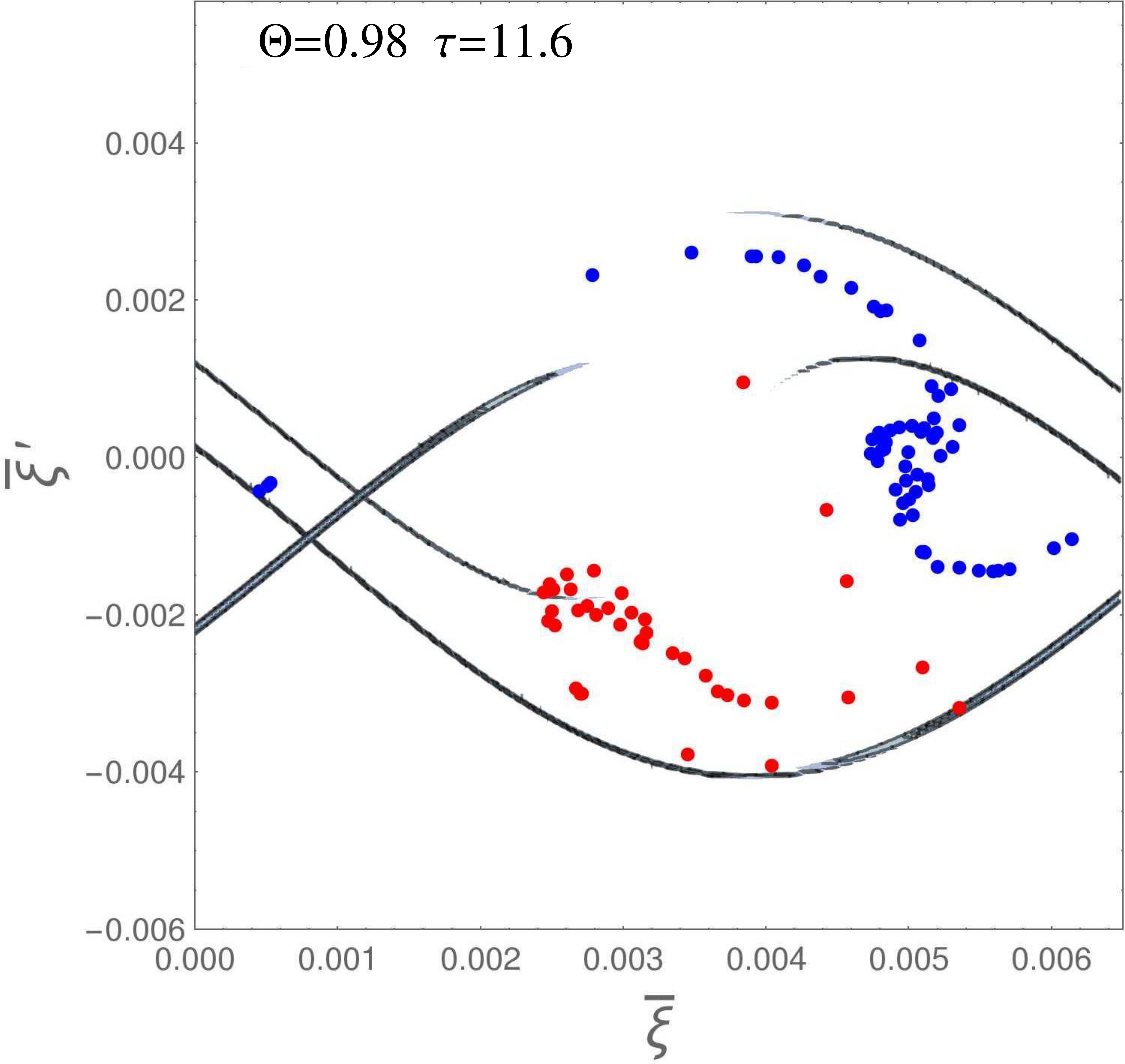}
\end{center}
\caption{Overlay of the phase-space snapshots (blue and red points for the beam 1 and 2, respectively) and the LCSs (thick lines) defined trough the contour plot peaks of the FTLE fields evaluated by \eref{eqftle}. The beam densities are set as $\su=\sd=0.5$ and we gave considered $\tau=11.6$ and $\deltat=5$. The different values of the beam initial velocity ratio ($\params$) are indicated in the graphs. (Color online)}
\label{fig:ftleTWM}
\end{figure}

Let us now analyze the transport processes in the two-beam system using the same methodology introduced above. In \figref{fig:ftleTWM}, we plot the LCS and the particle phase space at different values of $\params$ during the onset of the instability for $\tau=11.6$ and $\deltat=5$.

Considering the different colored particles initially belonging to different beams, it can be argued that they form a single phase-space structure for sufficiently small values of $1-\params$. In the first plot, two eye-shaped regions delimited by the LCS clearly emerge. Correspondingly, the particles of the different beams do not mix. The shape of the LCS becomes increasingly complicated as $\params\to1$. In the most general case, the LCSs do not divide the phase space into separate regions and, correspondingly, the particles of different beams can mix into a single structure. Therefore, the mixing process is greatly enhanced when LCSs belonging to different beams intersect; and they cannot be considered as finite time transport barriers between separate phase-space regions. We refer to $\params_{crit}$ as the smallest value of $\params$ compatible with the merging of the beams and give an estimate of such a threshold by increasing $\params$ until two LCS belonging to different beams intersect. From the plots in \figref{fig:ftleTWM}, we obtain $\params_{crit}$ between $0.96$ and $0.97$. For different beam densities, this method yields similar values of $\Theta_{crit}$. In the work \citep{BGP08}, a similar analysis was carried out in order to study the transition from local to global chaos in the magnetic field lines of a reconnecting current layer.

We conclude this Section with a derivation of $\Theta_{crit}$ using the Chirikov criterion.
Following \citep{OWM71}, the velocity spread of particles trapped by a single saturated beam mode is $\Delta v_j=\sqrt{2\phi_j}\,\etab$. The Chirikov parameter $S$ is, thus, defined by the ratio between the resonance width and the resonance separation of the two beams with initial velocity $v_1$ and $v_2$ (\ie $(v_{1}-v_{2})/v_{1}$). It can then be written as
\begin{equation}
S=\frac{2\Delta v_j}{(v_{1}-v_{2})/v_{1}}=2\frac{\sqrt{2 \phi_j}\, \bar{\eta}}{1- \Theta}\;.
\end{equation}
By the Chirikov criterion, two resonances overlap if $S>1$; therefore, in the saturation regime (\ie $\phi_j\simeq1$), we obtain $\Theta_{crit}\sim 0.97$, consistent with the previous analysis based on the LCSs (we remind that $\etab$ is fixed as $0.01$). We note that, even if the definition of LCSs we have used is not rigorous nor unique, their shape is \emph{a posteriori} consistent with particle dynamics and with the ensuing transport processes. The main advantage of using the LCS technique in studying transport processes between multiple resonant beams is that they can be calculated when the system is not near the saturated regime; thus, also when the Chirikov criterion cannot be straightforwardly applied.

\section{Concluding remarks}\label{conclu}

In this work, we have proposed to investigate the 1D bump-on-tail problem in terms of $n$ cold beams injected in a cold background plasma behaving as a linear dielectric medium; and $m\geq n$ self-consistently coupled nonlinear oscillators, mimicking the features of a warm fast particle distribution interacting with a spectrum of resonantly excited Langmuir waves. The use of the Hamiltonian formulation addressed in this work provides a convenient representation of the beam-plasma interaction and of ensuing transport processes at the disparate time and velocity scales underlying the complex behaviors of this system.

Specializing the model to two supra-thermal beams, we have determined the threshold ratio between the initial beam velocities, for which the two corresponding instabilities mutually interact instead of remaining separate growing modes. Such a value has been found to be $\sim0.96$ for the present simulation parameters, and is consistent with the estimates provided in \citep{L72} for a supra-thermal hot beam as well as with the Chirikov criterion. Above the threshold, the phase-space analysis of the particle motion clearly shows that transport appears as an exchange of particles between the two beams.

Furthermore, we have discussed the role of other quasi-degenerate modes, with particular attention to sideband modes. More specifically, we have shown that such additional modes may nonlinearly saturate at almost the same level reached by the most unstable waves. This fact suggests that the long time-scale evolution of the multi-beam system would be significantly affected by nonlinear sideband generation; and the same would hold for EP transport.

In the last par of the paper, we have investigated the onset of transport in the velocity space in terms of finite time transport barriers defined by the LCSs. The critical threshold for transport onset has been analyzed in terms of the LCSs of the system defined trough the FTLE field. The resulting critical value for the beam initial velocity ratio coincides with the Chirikov criterion and the direct inspection of particle motion in phase space; providing a very meaningful test of validity for the LCS technique in characterizing the phase-space dynamics.

Our analysis provides a qualitative and quantitative investigation framework for phase-space transport processes in the beam-plasma system. The advantage in using the present paradigm stands in the direct and very general characterization of the phase-space morphology. This leads to the definition of proper characteristic time and velocity scales related with spectral features and wave-particle resonances, together with the characterization of the diffusion coefficient in the quasi-linear limit expected for a large number of modes and a sufficiently broad spectrum.

\section*{Acknowledgement}
This work has been carried out within the framework of the EUROfusion Consortium and has received funding from the Euratom research and training programme 2014-2018 under grant agreement No 633053. The views and opinions expressed herein do not necessarily reflect those of the European Commission.

NC would like to thank Duccio Fanelli and Marco Pettini for their valuable advice on the code.

\bibliographystyle{jpp}
\bibliography{_PLASMA}

\begin{thebibliography}{91}
\expandafter\ifx\csname natexlab\endcsname\relax\def\natexlab#1{#1}\fi
\def\au#1{#1} \def\ed#1{#1} \def\yr#1{#1}\def\at#1{#1}\def\jt#1{\textit{#1}}
  \def\bt#1{#1}\def\bvol#1{\textbf{#1}} \def\vol#1{#1} \def\pg#1{#1}
  \def\publ#1{#1}\def\arxiv#1{#1}\def\org#1{#1}\def\st#1{\textit{#1}}

\bibitem[{Al'Tshul'} \& {Karpman}(1966)]{AK66}
{\sc \au{{Al'Tshul'}, L.M.} \& \au{{Karpman}, V.I.}} \yr{1966}  \at{Theory of
  nonlinear oscillations in a collisionless plasma}.  \jt{Sov. Phys. JEPT}
  \bvol{22}~(2),  \pg{361--369}.

\bibitem[{Antoni} {\em et~al.\/}(1998){Antoni}, {Elskens} \& {Escande}]{AEE98}
{\sc \au{{Antoni}, M.}, \au{{Elskens}, Y.} \& \au{{Escande}, D.F.}} \yr{1998}
  \at{Explicit reduction of n-body dynamics to self-consistent particle-wave
  interaction}.  \jt{Phys. Plasmas}  \bvol{5}~(4).

\bibitem[{Antoniazzi} {\em et~al.\/}(2005){Antoniazzi}, {DeNinno}, {Fanelli},
  {Guarino} \& {Ruffo}]{Duccio05}
{\sc \au{{Antoniazzi}, A.}, \au{{DeNinno}, G.}, \au{{Fanelli}, D.},
  \au{{Guarino}, A.} \& \au{{Ruffo}, S.}} \yr{2005}  \at{Wave-particle
  interaction: from plasma physics to the free-electron laser}.  \jt{J. Phys.
  Conf. Series}  \bvol{7},  \pg{143--153}.

\bibitem[{Antoniazzi} {\em et~al.\/}(2006){Antoniazzi}, {Elskens}, {Fanelli} \&
  {Ruffo}]{AEFR06}
{\sc \au{{Antoniazzi}, A.}, \au{{Elskens}, Y.}, \au{{Fanelli}, D.} \&
  \au{{Ruffo}, S.}} \yr{2006}  \at{Statistical mechanics and vlasov equation
  allow for a simplified hamiltonian description of single-pass free electron
  laser saturated dynamics}.  \jt{Eur. Phys. J. B}  \bvol{50}~(4),
  \pg{603--611}.

\bibitem[{Antoniazzi} {\em et~al.\/}(2008){Antoniazzi}, {Johal}, {Fanelli} \&
  {Ruffo}]{AJFR98}
{\sc \au{{Antoniazzi}, A.}, \au{{Johal}, R.~S.}, \au{{Fanelli}, D.} \&
  \au{{Ruffo}, S.}} \yr{2008}  \at{On the origin of quasi-stationary states in
  models of wave particle interaction}.  \jt{Comm. Nonlinear Sci Num. Sim.}
  \bvol{13},  \pg{2--10}.

\bibitem[{Berk} \& {Breizman}(1990{\natexlab{{\em a\/}}})]{BB90a}
{\sc \au{{Berk}, H.L.} \& \au{{Breizman}, B.N.}} \yr{1990{\natexlab{{\em
  a\/}}}}  \at{Saturation of a single mode driven by an energetic injected
  beam. i. plasma wave problem}.  \jt{Phys. Fluids B}  \bvol{2}~(9),
  \pg{2226--2234}.

\bibitem[{Berk} \& {Breizman}(1990{\natexlab{{\em b\/}}})]{BB90b}
{\sc \au{{Berk}, H.L.} \& \au{{Breizman}, B.N.}} \yr{1990{\natexlab{{\em
  b\/}}}}  \at{Saturation of a single mode driven by an energetic injected
  beam. ii. electrostatic ``universal'' destabilization mechanism}.  \jt{Phys.
  Fluids B}  \bvol{2}~(9),  \pg{2235--2245}.

\bibitem[{Berk} \& {Breizman}(1990{\natexlab{{\em c\/}}})]{BB90c}
{\sc \au{{Berk}, H.L.} \& \au{{Breizman}, B.N.}} \yr{1990{\natexlab{{\em
  c\/}}}}  \at{Saturation of a single mode driven by an energetic injected
  beam. iii. alfv{\'e}n wave problem}.  \jt{Phys. Fluids B}  \bvol{2}~(9),
  \pg{2246--2252}.

\bibitem[{Berk} {\em et~al.\/}(1996{\natexlab{{\em a\/}}}){Berk}, {Breizman},
  {Fitzpatrick}, {Pekker}, {Wong} \& {Wong}]{BB96b}
{\sc \au{{Berk}, H.L.}, \au{{Breizman}, B.N.}, \au{{Fitzpatrick}, J.},
  \au{{Pekker}, M.S.}, \au{{Wong}, H.V.} \& \au{{Wong}, K.L.}}
  \yr{1996{\natexlab{{\em a\/}}}}  \at{Nonlinear response of driven systems in
  weak turbulence theory}.  \jt{Phys. Plasmas}  \bvol{3}~(5),  \pg{1827--1838}.

\bibitem[{Berk} {\em et~al.\/}(1995{\natexlab{{\em a\/}}}){Berk}, {Breizman},
  {Fitzpatrick} \& {Wong}]{BB95b}
{\sc \au{{Berk}, H.L.}, \au{{Breizman}, B.N.}, \au{{Fitzpatrick}, J.} \&
  \au{{Wong}, H.V.}} \yr{1995{\natexlab{{\em a\/}}}}  \at{Line broadened
  quasi-linear burst model [fusion plasma]}.  \jt{Nucl. Fusion}
  \bvol{35}~(12),  \pg{1661--1668}.

\bibitem[{Berk} {\em et~al.\/}(1995{\natexlab{{\em b\/}}}){Berk}, {Breizman} \&
  {Pekker}]{BB95a}
{\sc \au{{Berk}, H.L.}, \au{{Breizman}, B.N.} \& \au{{Pekker}, M.}}
  \yr{1995{\natexlab{{\em b\/}}}}  \at{Numerical simulation of bump-on-tail
  instability with source and sink}.  \jt{Phys. Plasmas}  \bvol{2}~(8),
  \pg{3007--1259}.

\bibitem[{Berk} {\em et~al.\/}(1996{\natexlab{{\em b\/}}}){Berk}, {Breizman} \&
  {Pekker}]{BB96a}
{\sc \au{{Berk}, H.L.}, \au{{Breizman}, B.N.} \& \au{{Pekker}, M.}}
  \yr{1996{\natexlab{{\em b\/}}}}  \at{Nonlinear dynamics of a driven mode near
  marginal stability}.  \jt{Phys. Rev. Lett.}  \bvol{76},  \pg{1256}.

\bibitem[{Berk} {\em et~al.\/}(1997){Berk}, {Breizman} \& {Pekker}]{BB97a}
{\sc \au{{Berk}, H.L.}, \au{{Breizman}, B.N.} \& \au{{Pekker}, M.S.}} \yr{1997}
   \at{Nonlinear theory of kinetic instabilities near threshold}.  \jt{Plasma
  Phys. Rept.}  \bvol{23}~(9),  \pg{778--788}.

\bibitem[{Borgogno} {\em et~al.\/}(2008){Borgogno}, {Grasso}, {Pegoraro} \&
  {Schep}]{BGP08}
{\sc \au{{Borgogno}, D.}, \au{{Grasso}, D.}, \au{{Pegoraro}, F.} \&
  \au{{Schep}, T.J.}} \yr{2008}  \at{Stable and unstable invariant manifolds in
  a partially chaotic magnetic configuration generated by nonlinear
  reconnection}.  \jt{Phys. Plasmas}  \bvol{15}~(10),  \pg{102308}.

\bibitem[{Borgogno} {\em et~al.\/}(2011{\natexlab{{\em a\/}}}){Borgogno},
  {Grasso}, {Pegoraro} \& {Schep}]{Bo11a}
{\sc \au{{Borgogno}, D.}, \au{{Grasso}, D.}, \au{{Pegoraro}, F.} \&
  \au{{Schep}, T.J.}} \yr{2011{\natexlab{{\em a\/}}}}  \at{Barriers in the
  transition to global chaos in collisionless magnetic reconnection. i. ridges
  of the finite time lyapunov exponent field}.  \jt{Phys. Plasmas}
  \bvol{18}~(10),  \pg{102307}.

\bibitem[{Borgogno} {\em et~al.\/}(2011{\natexlab{{\em b\/}}}){Borgogno},
  {Grasso}, {Pegoraro} \& {Schep}]{Bo11b}
{\sc \au{{Borgogno}, D.}, \au{{Grasso}, D.}, \au{{Pegoraro}, F.} \&
  \au{{Schep}, T.J.}} \yr{2011{\natexlab{{\em b\/}}}}  \at{Barriers in the
  transition to global chaos in collisionless magnetic reconnection. ii. field
  line spectroscopy}.  \jt{Phys. Plasmas}  \bvol{18}~(10),  \pg{102308}.

\bibitem[{Breizman}(2011)]{Br11}
{\sc \au{{Breizman}, B.}} \yr{2011}  \at{Nonlinear consequences of energetic
  particle instabilities}.  \jt{Fus. Sci. Technol.}  \bvol{59}~(3),
  \pg{549--560}.

\bibitem[{Breizman} {\em et~al.\/}(1997){Breizman}, {Berk}, {Pekker},
  {Porcelli}, {Stupakov} \& {Wong}]{BB97b}
{\sc \au{{Breizman}, B.N.}, \au{{Berk}, H.L.}, \au{{Pekker}, M.S.},
  \au{{Porcelli}, F.}, \au{{Stupakov}, G.V.} \& \au{{Wong}, K.L.}} \yr{1997}
  \at{Critical nonlinear phenomena for kinetic instabilities near threshold}.
  \jt{Phys. Plasmas}  \bvol{4}~(5),  \pg{1559--1568}.

\bibitem[{Breizman} {\em et~al.\/}(1993){Breizman}, {Berk} \& {Ye}]{BB93}
{\sc \au{{Breizman}, B.N.}, \au{{Berk}, H.L.} \& \au{{Ye}, H.}} \yr{1993}
  \at{Collective transport of alpha particles due to alfv{\'e}n wave
  instability}.  \jt{Phys. Fluids B}  \bvol{5}~(9),  \pg{3217--3226}.

\bibitem[{Breizman} \& {Sharapov}(2011)]{BS11}
{\sc \au{{Breizman}, B.N.} \& \au{{Sharapov}, S.E.}} \yr{2011}  \at{Major
  minority: energetic particles in fusion plasmas}.  \jt{Plasma Phys. Contr.
  Fusion}  \bvol{53}~(5),  \pg{054001}.

\bibitem[{Briguglio} {\em et~al.\/}(2014){Briguglio}, {Wang}, {Zonca}, {Vlad},
  {Fogaccia}, {Di Troia} \& {Fusco}]{BW14pop}
{\sc \au{{Briguglio}, S.}, \au{{Wang}, X.}, \au{{Zonca}, F.}, \au{{Vlad}, G.},
  \au{{Fogaccia}, G.}, \au{{Di Troia}, C.} \& \au{{Fusco}, V.}} \yr{2014}
  \at{Analysis of the nonlinear behavior of shear-alfv{\'e}n modes in tokamaks
  based on hamiltonian mapping techniques}.  \jt{Phys. Plasmas}
  \bvol{21}~(11),  \pg{112301}.

\bibitem[{Carlevaro} {\em et~al.\/}(2014){Carlevaro}, {Fanelli}, {Garbet},
  {Ghendrih}, {Montani} \& {Pettini}]{CFGGMP14}
{\sc \au{{Carlevaro}, N.}, \au{{Fanelli}, D.}, \au{{Garbet}, X.},
  \au{{Ghendrih}, P.}, \au{{Montani}, G.} \& \au{{Pettini}, M.}} \yr{2014}
  \at{Beam-plasma instability and fast particles: the lynden-bell approach}.
  \jt{Plasma Phys. Contr. Fusion}  \bvol{56}~(3),  \pg{035013}.

\bibitem[{Chen} \& {Zonca}(2007)]{CZ07}
{\sc \au{{Chen}, L.} \& \au{{Zonca}, F.}} \yr{2007}  \at{Theory of alfv{\'e}n
  waves and energetic particle physics in burning plasmas}.  \jt{Nucl. Fusion}
  \bvol{47}~(10),  \pg{S727--S734}.

\bibitem[{Chen} \& {Zonca}(2013)]{CZ13}
{\sc \au{{Chen}, L.} \& \au{{Zonca}, F.}} \yr{2013}  \at{On nonlinear physics
  of shear alfv{\'e}n waves}.  \jt{Phys. Plasmas}  \bvol{20}~(5),  \pg{055402}.

\bibitem[{Chen} \& {Zonca}(2015)]{ZCrmp}
{\sc \au{{Chen}, L.} \& \au{{Zonca}, F.}} \yr{2015}  \at{Physics of alfv\'en
  waves and energetic particles in burning plasmas}.  \jt{submitted to Rev.
  Mod. Phys.} .

\bibitem[{Chian} {\em et~al.\/}(2014){Chian}, {Rempel}, {Aulanier},
  {Schmieder}, {Shadden}, {Welsch} \& {Yeates}]{CRASSWY14}
{\sc \au{{Chian}, A.}, \au{{Rempel}, E.L.}, \au{{Aulanier}, G.},
  \au{{Schmieder}, B.}, \au{{Shadden}, S.C.}, \au{{Welsch}, B.T.} \&
  \au{{Yeates}, A.R.}} \yr{2014}  \at{Detection of coherent structures in
  photospheric turbulent flows}.  \jt{ApJ}  \bvol{786}~(1),  \pg{51}.

\bibitem[{Chirikov}(1979)]{Ch79}
{\sc \au{{Chirikov}, B.V.}} \yr{1979}  \at{A universal instability of
  many-dimensional oscillator systems}.  \jt{Phys. Rept.}  \bvol{52}~(5),
  \pg{263--379}.

\bibitem[{Deneef}(1975)]{DeN75}
{\sc \au{{Deneef}, P.}} \yr{1975}  \at{Two waves on a beam-plasma system}.
  \jt{Phys. Fluids}  \bvol{18},  \pg{1209--1212}.

\bibitem[{Elskens} \& {Escande}(2003)]{EEbook}
{\sc \au{{Elskens}, Y.} \& \au{{Escande}, D.F.}} \yr{2003} {\em Microscopic
  Dynamics of Plasmas Chaos\/}.  \publ{Taylor Francis Ltd}.

\bibitem[{Esarey} {\em et~al.\/}(1996){Esarey}, {Sprangle}, {Krall} \&
  {Ting}]{ES96}
{\sc \au{{Esarey}, E.}, \au{{Sprangle}, P.}, \au{{Krall}, J.} \& \au{{Ting},
  A.}} \yr{1996}  \at{Overview of plasma-based accelerator concepts}.  \jt{IEEE
  Tran. Plasma Science}  \bvol{24}~(2),  \pg{252--288}.

\bibitem[{Escande} \& {Elskens}(2008)]{EE08}
{\sc \au{{Escande}, D.F.} \& \au{{Elskens}, Y.}} \yr{2008}
  \at{Self-consistency vanishes in the plateau regime of the bump-on-tail
  instability}.  \jt{arXiv:0807.1839} .

\bibitem[{Evstatiev} {\em et~al.\/}(2003){Evstatiev}, {Horton} \&
  {Morrison}]{EHM03}
{\sc \au{{Evstatiev}, E.G.}, \au{{Horton}, W.} \& \au{{Morrison}, P.J.}}
  \yr{2003}  \at{Multiwave model for plasma-wave interaction}.  \jt{Phys.
  Plasmas}  \bvol{10}~(10),  \pg{4090--4094}.

\bibitem[{Falessi} {\em et~al.\/}(2015){Falessi}, {Pegoraro} \& {Schep}]{MPJPP}
{\sc \au{{Falessi}, M.V.}, \au{{Pegoraro}, F.} \& \au{{Schep}, T.J.}} \yr{2015}
   \at{Geometric structures, lobe dynamics, and lagrangian transport in flows
  with aperiodic time-dependence, with applications to rossby wave flow}.
  \jt{J. Plasma Phys}  \bvol{81}~(5),  \pg{495810505}.

\bibitem[{Farina} {\em et~al.\/}(1994){Farina}, {Casagrande}, {Colombo} \&
  {Pozzoli}]{Farina94}
{\sc \au{{Farina}, D.}, \au{{Casagrande}, F.}, \au{{Colombo}, U.} \&
  \au{{Pozzoli}, R.}} \yr{1994}  \at{Hamiltonian analysis of the transition to
  the high-gain regime in a compton free-electron-laser amplifier}.  \jt{Phys.
  Rev. E}  \bvol{49}~(2),  \pg{1603--1609}.

\bibitem[{Farina} \& {Pozzoli}(2004)]{Farina04}
{\sc \au{{Farina}, D.} \& \au{{Pozzoli}, R.}} \yr{2004}  \at{Large-amplitude
  oscillations and chaos in a hamiltonian plasma system with many degrees of
  freedom}.  \jt{Phys. Rev. E}  \bvol{70}~(3),  \pg{036407}.

\bibitem[Fasoli {\em et~al.\/}(2007)Fasoli, Gormenzano, Berk, Breizman,
  Briguglio, Darrow, Gorelenkov, Heidbrink, Jaun, Konovalov, Nazikian,
  Noterdaeme, Sharapov, Shinohara, Testa, Tobita, Todo, Vlad \& Zonca]{Fa07}
{\sc \au{Fasoli, A.}, \au{Gormenzano, C.}, \au{Berk, H.L.}, \au{Breizman, B.},
  \au{Briguglio, S.}, \au{Darrow, D.S.}, \au{Gorelenkov, N.}, \au{Heidbrink,
  W.W.}, \au{Jaun, A.}, \au{Konovalov, S.V.}, \au{Nazikian, R.},
  \au{Noterdaeme, J.-M.}, \au{Sharapov, S.}, \au{Shinohara, K.}, \au{Testa,
  D.}, \au{Tobita, K.}, \au{Todo, Y.}, \au{Vlad, G.} \& \au{Zonca, F.}}
  \yr{2007}  \at{Chapter 5: Physics of energetic ions}.  \jt{Nucl. Fusion}
  \bvol{47}~(6),  \pg{S264--S284}.

\bibitem[{Firpo} \& {Elskens}(2000)]{FE00}
{\sc \au{{Firpo}, M.-C.} \& \au{{Elskens}, Y.}} \yr{2000}  \at{Phase transition
  in the collisionless damping regime for wave-particle interaction}.
  \jt{Phys. Rev. Lett.}  \bvol{84}~(15),  \pg{3318--3321}.

\bibitem[{Garth} {\em et~al.\/}(2007){Garth}, {Gerhardt}, {Tricoche} \&
  {Hagen}]{Ga07}
{\sc \au{{Garth}, C.}, \au{{Gerhardt}, F.}, \au{{Tricoche}, X.} \& \au{{Hagen},
  H.}} \yr{2007}  \at{Efficient computation and visualization of coherent
  structures in fluid flow applications}.  \jt{IEEE Transactions on
  Visualization and Computer Graphics}  \bvol{13}~(6),  \pg{1464--1471}.

\bibitem[{Ghantous} {\em et~al.\/}(2014){Ghantous}, {Berk} \&
  {Gorelenkov}]{GBG14}
{\sc \au{{Ghantous}, K.}, \au{{Berk}, H.L.} \& \au{{Gorelenkov}, N.N.}}
  \yr{2014}  \at{Comparing the line broadened quasilinear model to vlasov
  code}.  \jt{Phys. Plasmas}  \bvol{21}~(3),  \pg{032119}.

\bibitem[{Ghantous} {\em et~al.\/}(2012){Ghantous}, {Gorelenkov}, {Berk},
  {Heidbrink} \& {Van Zeeland}]{GG12}
{\sc \au{{Ghantous}, K.}, \au{{Gorelenkov}, N.N.}, \au{{Berk}, H.L.},
  \au{{Heidbrink}, W.W.} \& \au{{Van Zeeland}, M.A.}} \yr{2012}  \at{1.5d
  quasilinear model and its application on beams interacting with alfv{\'e}n
  eigenmodes in diii-d}.  \jt{Phys. Plasmas}  \bvol{19}~(9),  \pg{092511}.

\bibitem[{Gorelenkov} {\em et~al.\/}(2014){Gorelenkov}, {Pinches} \&
  {Toi}]{GPT14}
{\sc \au{{Gorelenkov}, N.N.}, \au{{Pinches}, S.D.} \& \au{{Toi}, K.}} \yr{2014}
   \at{Energetic particle physics in fusion research in preparation for burning
  plasma experiments}.  \jt{Nucl. Fusion}  \bvol{54}~(12),  \pg{125001}.

\bibitem[{Haller}(2001)]{H01}
{\sc \au{{Haller}, G.}} \yr{2001}  \at{Distinguished material surfaces and
  coherent structures in three-dimensional fluid flows}.  \jt{Physica D}
  \bvol{149}~(4),  \pg{248--277}.

\bibitem[{Haller}(2011)]{H11}
{\sc \au{{Haller}, G.}} \yr{2011}  \at{A variational theory of hyperbolic
  lagrangian coherent structures}.  \jt{Physica D}  \bvol{240}~(7),  \pg{574}.

\bibitem[{Haller}(2015)]{Haller15l}
{\sc \au{{Haller}, G.}} \yr{2015}  \at{Lagrangian coherent structures}.
  \jt{Annual Rev. Fluid Mech.}  \bvol{47}~(1),  \pg{137--162}.

\bibitem[{Heidbrink}(2008)]{H08}
{\sc \au{{Heidbrink}, W.W.}} \yr{2008}  \at{Basic physics of alfv{\'e}n
  instabilities driven by energetic particles in toroidally confined plasmas}.
  \jt{Phys. Plasmas}  \bvol{15}~(5),  \pg{055501}.

\bibitem[{Kasten} {\em et~al.\/}(2010){Kasten}, {Petz}, {Hotz}, {Hege}, {Noack}
  \& {Tadmor}]{KPHHNT10}
{\sc \au{{Kasten}, J.}, \au{{Petz}, C.}, \au{{Hotz}, I.}, \au{{Hege}, H.-C.},
  \au{{Noack}, B.R.} \& \au{{Tadmor}, G.}} \yr{2010}  \at{Lagrangian feature
  extraction of the cylinder wake}.  \jt{Phys. Fluids}  \bvol{22}~(9),
  \pg{091108}.

\bibitem[{Keinigs} \& {Jones}(1987)]{KJ86}
{\sc \au{{Keinigs}, R.} \& \au{{Jones}, M.E.}} \yr{1987}  \at{Two-dimensional
  dynamics of the plasma wakefield accelerator}.  \jt{Phys. Fluids}
  \bvol{30}~(1),  \pg{252--263}.

\bibitem[{Klimontovich}(1967)]{Klim}
{\sc \au{{Klimontovich}, Yu.L.}} \yr{1967} {\em The statistical theory of
  Non-equilibrium processes in a plasma\/}.  \publ{MIT Press}.

\bibitem[{Krafft} \& {Volokitin}(2014)]{KV14}
{\sc \au{{Krafft}, C.} \& \au{{Volokitin}, A.}} \yr{2014}  \at{Hamiltonian
  models for resonant wave-particle interaction processes in magnetized and
  inhomogeneous plasmas}.  \jt{Eur. Phys. J. D}  \bvol{68}~(12),  \pg{370}.

\bibitem[{Lacina} {\em et~al.\/}(1976){Lacina}, {Krl\'in} \&
  {K$\ddot{\textrm{o}}$rbel}]{LKK75}
{\sc \au{{Lacina}, J.}, \au{{Krl\'in}, L.} \& \au{{K$\ddot{\textrm{o}}$rbel},
  S.}} \yr{1976}  \at{Effect of beam density and of higher harmonics on
  beam-plasma interaction}.  \jt{Plasma Phys.}  \bvol{18},  \pg{471--483}.

\bibitem[{Lauber}(2013)]{La13}
{\sc \au{{Lauber}, P.}} \yr{2013}  \at{Super-thermal particles in hot plasmas -
  kinetic models, numerical solution strategies, and comparison to tokamak
  experiments}.  \jt{Phys. Rept.}  \bvol{533}~(2),  \pg{33--68}.

\bibitem[{Lekien} \& {Ross}(2010)]{LR10}
{\sc \au{{Lekien}, F.} \& \au{{Ross}, S.D.}} \yr{2010}  \at{The computation of
  finite-time lyapunov exponents on unstructured meshes and for non-euclidean
  manifolds}.  \jt{Chaos}  \bvol{20}~(1),  \pg{017505}.

\bibitem[{Leoncini} \& {Zaslavsky}(2002)]{LZ02}
{\sc \au{{Leoncini}, X.} \& \au{{Zaslavsky}, G.M.}} \yr{2002}  \at{Jets,
  stickiness, and anomalous transport}.  \jt{Phys. Rev. E}  \bvol{65}~(4),
  \pg{046216}.

\bibitem[{Lesur} \& {Idomura}(2012)]{LI12}
{\sc \au{{Lesur}, M.} \& \au{{Idomura}, Y.}} \yr{2012}  \at{Nonlinear
  categorization of the energetic-beam-driven instability with drag and
  diffusion}.  \jt{Nucl. Fusion}  \bvol{52}~(9),  \pg{094004}.

\bibitem[{Lesur} {\em et~al.\/}(2009){Lesur}, {Idomura} \& {Garbet}]{LI09}
{\sc \au{{Lesur}, M.}, \au{{Idomura}, Y.} \& \au{{Garbet}, X.}} \yr{2009}
  \at{Fully nonlinear features of the energetic beam-driven instability}.
  \jt{Phys. Plasmas}  \bvol{16}~(9),  \pg{092305}.

\bibitem[{Lesur} {\em et~al.\/}(2010){Lesur}, {Idomura}, {Shinohara}, {Garbet}
  \& {JT-60 Team}]{LI10}
{\sc \au{{Lesur}, M.}, \au{{Idomura}, Y.}, \au{{Shinohara}, K.}, \au{{Garbet},
  X.} \& \au{{JT-60 Team}}} \yr{2010}  \at{Spectroscopic determination of
  kinetic parameters for frequency sweeping alfv{\'e}n eigenmodes}.  \jt{Phys.
  Plasmas}  \bvol{17}~(12),  \pg{122311}.

\bibitem[{Levin} {\em et~al.\/}(1972){Levin}, {Lyubarski{\v i}}, {Onishchenko},
  {Shapiro} \& {Shevchenko}]{L72}
{\sc \au{{Levin}, M.B.}, \au{{Lyubarski{\v i}}, M.G.}, \au{{Onishchenko},
  I.N.}, \au{{Shapiro}, V.D.} \& \au{{Shevchenko}, V.I.}} \yr{1972}
  \at{Contribution to the nonlinear theory of kinetic instability of an
  electron beam in plasma}.  \jt{Sov. Phys. JEPT}  \bvol{35}~(5),
  \pg{898--901}.

\bibitem[Lifshitz \& {Pitaevskii}(1976)]{LP81}
{\sc \au{Lifshitz, E.M.} \& \au{{Pitaevskii}, L.P.}} \yr{1976} {\em Course of
  Theoretical Physics, Volume 10: Physical Kinetics\/}.
  \publ{Butterworth-Heinemann}.

\bibitem[{Lilley} \& {Breizman}(2012)]{LB12}
{\sc \au{{Lilley}, M.K.} \& \au{{Breizman}, B.N.}} \yr{2012}  \at{Convective
  transport of fast particles in dissipative plasmas near an instability
  threshold}.  \jt{Nucl. Fusion}  \bvol{52}~(9),  \pg{094002}.

\bibitem[{Lilley} \& {Nyqvist}(2014)]{LN14}
{\sc \au{{Lilley}, M.K.} \& \au{{Nyqvist}, R.M.}} \yr{2014}  \at{Formation of
  phase space holes and clumps}.  \jt{Phys. Rev. Lett.}  \bvol{112}~(15),
  \pg{155002}.

\bibitem[{Litos} {\em et~al.\/}(2014){Litos}, {Adli}, {An}, {Clarke},
  {Clayton}, {Corde}, {Delahaye}, {England}, {Fisher}, {Frederico}, {Gessner},
  {Green}, {Hogan}, {Joshi}, {Lu}, {Marsh}, {Mori}, {Muggli},
  {Vafaei-Najafabadi}, {Walz}, {White}, {Wu}, {Yakimenko} \& {Yocky}]{Li14}
{\sc \au{{Litos}, M.}, \au{{Adli}, E.}, \au{{An}, W.}, \au{{Clarke}, C.I.},
  \au{{Clayton}, C.E.}, \au{{Corde}, S.}, \au{{Delahaye}, J.P.}, \au{{England},
  R.J.}, \au{{Fisher}, A.S.}, \au{{Frederico}, J.}, \au{{Gessner}, S.},
  \au{{Green}, S.Z.}, \au{{Hogan}, M.J.}, \au{{Joshi}, C.}, \au{{Lu}, W.},
  \au{{Marsh}, K.A.}, \au{{Mori}, W.B.}, \au{{Muggli}, P.},
  \au{{Vafaei-Najafabadi}, N.}, \au{{Walz}, D.}, \au{{White}, G.}, \au{{Wu},
  Z.}, \au{{Yakimenko}, V.} \& \au{{Yocky}, G.}} \yr{2014}  \at{High-efficiency
  acceleration of an electron beam in a plasma wakefield accelerator}.
  \jt{Nature}  \bvol{515}~(7525),  \pg{92--95}.

\bibitem[{Malhotra} \& {Wiggins}(1998)]{MW98}
{\sc \au{{Malhotra}, N.} \& \au{{Wiggins}, S.}} \yr{1998}  \at{Geometric
  structures, lobe dynamics, and lagrangian transport in flows with aperiodic
  time-dependence, with applications to rossby wave flow}.  \jt{J. Nonlinear
  Sci.}  \bvol{8}~(4),  \pg{401--456}.

\bibitem[{Matsiborko} {\em et~al.\/}(1972){Matsiborko}, {Onishchenko},
  {Shapiro} \& {Shevchenko}]{Ma72}
{\sc \au{{Matsiborko}, N.G.}, \au{{Onishchenko}, I.N.}, \au{{Shapiro}, V.D.} \&
  \au{{Shevchenko}, V.I.}} \yr{1972}  \at{On non-linear theory of instability
  of a mono-energetic electron beam in plasma}.  \jt{Plasma Phys.}
  \bvol{14}~(6),  \pg{591--600}.

\bibitem[{Mynick} \& {Kaufman}(1978)]{MK78}
{\sc \au{{Mynick}, H.E.} \& \au{{Kaufman}, A.N.}} \yr{1978}  \at{Soluble theory
  of nonlinear beam-plasma interaction}.  \jt{Phys. Fluids}  \bvol{21},
  \pg{653--663}.

\bibitem[{O'Neil} \& {Malmberg}(1968)]{OM68}
{\sc \au{{O'Neil}, T.M.} \& \au{{Malmberg}, J.H.}} \yr{1968}  \at{Transition of
  the dispersion roots from beam-type to landau-type solutions}.  \jt{Phys.
  Fluids}  \bvol{11}~(8),  \pg{1754--1760}.

\bibitem[{O'Neil} {\em et~al.\/}(1971){O'Neil}, {Winfrey} \& {Malmberg}]{OWM71}
{\sc \au{{O'Neil}, T.M.}, \au{{Winfrey}, J.H.} \& \au{{Malmberg}, J.H.}}
  \yr{1971}  \at{Nonlinear interaction of a small cold beam and a plasma}.
  \jt{Phys. Fluids}  \bvol{14}~(6),  \pg{1204--1212}.

\bibitem[{Onishchenko} {\em et~al.\/}(1970){Onishchenko}, {Linetskii},
  {Matsiborko}, {Shapiro} \& {Shevchenko}]{OL70}
{\sc \au{{Onishchenko}, I.N.}, \au{{Linetskii}, A.R.}, \au{{Matsiborko}, N.G.},
  \au{{Shapiro}, V.D.} \& \au{{Shevchenko}, V.I.}} \yr{1970}  \at{Contribution
  to the nonlinear theory of excitation of a monochromatic plasma wave by an
  electron beam}.  \jt{JETP Letters}  \bvol{12}~(8).

\bibitem[{Pascucci} {\em et~al.\/}(2010){Pascucci}, {Tricoche}, {Hagen} \&
  {Tierny}]{Sa12}
{\sc \au{{Pascucci}, V.}, \au{{Tricoche}, X.}, \au{{Hagen}, H.} \&
  \au{{Tierny}, J.}} \yr{2010} {\em Topological Methods in Data Analysis and
  Visualization: theory, algorithms, and applications\/}.  \publ{Springer}.

\bibitem[{Peacock} \& {Haller}(2013)]{PH13}
{\sc \au{{Peacock}, T.} \& \au{{Haller}, G.}} \yr{2013}  \at{Lagrangian
  coherent structures:: The hidden skeleton of fluid flows}.  \jt{Phys. Today}
  \bvol{66}~(2),  \pg{41--47}.

\bibitem[{Pechhacker} \& {Tsiklauri}(2014)]{PIC3D}
{\sc \au{{Pechhacker}, R.} \& \au{{Tsiklauri}, D.}} \yr{2014}
  \at{Three-dimensional particle-in-cell simulation of electron acceleration by
  langmuir waves in an inhomogeneous plasma}.  \jt{Phys. Plasmas}
  \bvol{21}~(1),  \pg{012903}.

\bibitem[{Pinches} {\em et~al.\/}(2015){Pinches}, {Chapman}, {Lauber},
  {Oliver}, {Sharapov}, {Shinohara} \& {Tani}]{Pi15}
{\sc \au{{Pinches}, S.D.}, \au{{Chapman}, I.T.}, \au{{Lauber}, P.W.},
  \au{{Oliver}, H.J.C.}, \au{{Sharapov}, S.E.}, \au{{Shinohara}, K.} \&
  \au{{Tani}, K.}} \yr{2015}  \at{Energetic ions in iter plasmas}.  \jt{Phys.
  Plasmas}  \bvol{22}~(2),  \pg{021807}.

\bibitem[{Schneller} {\em et~al.\/}(2013){Schneller}, {Lauber}, {Bilato},
  {Garc{\'{\i}}a-Mu{\~n}oz}, {Br{\"u}dgam}, {G{\"u}nter} \& {the ASDEX Upgrade
  Team}]{ML13}
{\sc \au{{Schneller}, M.}, \au{{Lauber}, P.}, \au{{Bilato}, R.},
  \au{{Garc{\'{\i}}a-Mu{\~n}oz}, M.}, \au{{Br{\"u}dgam}, M.}, \au{{G{\"u}nter},
  S.} \& \au{{the ASDEX Upgrade Team}}} \yr{2013}  \at{Multi-mode alfv{\'e}nic
  fast particle transport and losses: numerical versus experimental
  observation}.  \jt{Nucl. Fusion}  \bvol{53}~(12),  \pg{123003}.

\bibitem[{Schneller} {\em et~al.\/}(2012){Schneller}, {Lauber}, {Br{\"u}dgam},
  {Pinches} \& {G{\"u}nter}]{ML12}
{\sc \au{{Schneller}, M.}, \au{{Lauber}, P.}, \au{{Br{\"u}dgam}, M.},
  \au{{Pinches}, S.D.} \& \au{{G{\"u}nter}, S.}} \yr{2012}  \at{Double-resonant
  fast particle-wave interaction}.  \jt{Nucl. Fusion}  \bvol{52}~(10),
  \pg{103019}.

\bibitem[{Senatore} \& {Ross}(2011)]{Se11}
{\sc \au{{Senatore}, C.} \& \au{{Ross}, S.D.}} \yr{2011}  \at{Detection and
  characterization of transport barriers in complex flows via ridge extraction
  of the finite time lyapunov exponent field}.  \jt{I. J. Num. Methods Eng.}
  \bvol{86}~(9),  \pg{1163--1174}.

\bibitem[{Shadden} {\em et~al.\/}(2005){Shadden}, {Lekien} \& {Marsden}]{SLM05}
{\sc \au{{Shadden}, S.C.}, \au{{Lekien}, F.} \& \au{{Marsden}, J.E.}} \yr{2005}
   \at{Definition and properties of lagrangian coherent structures from
  finite-time lyapunov exponents in two-dimensional aperiodic flows}.
  \jt{Physica D}  \bvol{212}~(3-4),  \pg{271--304}.

\bibitem[{Shapiro}(1963)]{Sh63}
{\sc \au{{Shapiro}, V.D.}} \yr{1963}  \at{Nonlinear theory of the interaction
  of a monoenergetic beam with a plasma}.  \jt{Sov. Phys. JEPT}  \bvol{17}~(2),
   \pg{416--423}.

\bibitem[{Shapiro} \& {Shevchenko}(1971)]{SS71}
{\sc \au{{Shapiro}, V.D.} \& \au{{Shevchenko}, V.I.}} \yr{1971}
  \at{Contribution to the nonlinear theory of relaxation of a ``monoenergetic''
  beam in a plasma}.  \jt{Sov. Phys. JEPT}  \bvol{33}~(3),  \pg{555--561}.

\bibitem[{Shoucri}(2010)]{Sho10}
{\sc \au{{Shoucri}, M.}} \yr{2010} {\em Eulerian Codes for the Numerical
  Solution of the Kinetic Equations of Plasmas\/}.  \publ{Nova Science Pub
  Inc}.

\bibitem[{Smith} \& {Pereira}(1978)]{SP78}
{\sc \au{{Smith}, G.R.} \& \au{{Pereira}, N.R.}} \yr{1978}  \at{Phase-locked
  particle motion in a large-amplitude plasma wave}.  \jt{Phys. Fluids}
  \bvol{21}~(12),  \pg{2253--2262}.

\bibitem[{Tang} {\em et~al.\/}(2010){Tang}, {Mathur}, {Haller}, {Hahn} \&
  {Ruggiero}]{MHPRS07}
{\sc \au{{Tang}, W.}, \au{{Mathur}, M.}, \au{{Haller}, G.}, \au{{Hahn}, D.C.}
  \& \au{{Ruggiero}, F.H.}} \yr{2010}  \at{Lagrangian coherent structures near
  a subtropical jet stream}.  \jt{J. Atmospheric Sci.}  \bvol{67}~(7),
  \pg{2307--2319}.

\bibitem[{Tennyson} {\em et~al.\/}(1994){Tennyson}, {Meiss} \&
  {Morrison}]{TMM94}
{\sc \au{{Tennyson}, J.L.}, \au{{Meiss}, J.D.} \& \au{{Morrison}, P.J.}}
  \yr{1994}  \at{Self-consistent chaos in the beam-plasma instability}.
  \jt{Physica D}  \bvol{71}~(1-2),  \pg{1--17}.

\bibitem[{Thompson}(1971)]{Th71}
{\sc \au{{Thompson}, J.R.}} \yr{1971}  \at{Nonlinear evolution of collisionless
  electron beam-plasma systems}.  \jt{Phys. Fluids}  \bvol{14}~(7),
  \pg{1532--1541}.

\bibitem[{Vlad} {\em et~al.\/}(2013){Vlad}, {Briguglio}, {Fogaccia}, {Zonca},
  {Fusco} \& {Wang}]{VB13}
{\sc \au{{Vlad}, G.}, \au{{Briguglio}, S.}, \au{{Fogaccia}, G.}, \au{{Zonca},
  F.}, \au{{Fusco}, V.} \& \au{{Wang}, X.}} \yr{2013}  \at{Electron fishbone
  simulations in tokamak equilibria using xhmgc}.  \jt{Nucl. Fusion}
  \bvol{53}~(8),  \pg{083008}.

\bibitem[{Volokitin} \& {Krafft}(2012)]{VK12}
{\sc \au{{Volokitin}, A.} \& \au{{Krafft}, C.}} \yr{2012}  \at{Velocity
  diffusion in plasma waves excited by electron beams}.  \jt{Plasma Phys.
  Control. Fusion}  \bvol{54}~(8),  \pg{085002}.

\bibitem[{Voth} {\em et~al.\/}(2002){Voth}, {Haller} \& {Gollub}]{VHG02}
{\sc \au{{Voth}, G.A.}, \au{{Haller}, G.} \& \au{{Gollub}, J.P.}} \yr{2002}
  \at{Experimental measurements of stretching fields in fluid mixing}.
  \jt{Phys. Rev. Lett.}  \bvol{88}~(25),  \pg{254501}.

\bibitem[{Wang} {\em et~al.\/}(2011){Wang}, {Briguglio}, {Chen}, {Di Troia},
  {Fogaccia}, {Vlad} \& {Zonca}]{WB11}
{\sc \au{{Wang}, X.}, \au{{Briguglio}, S.}, \au{{Chen}, L.}, \au{{Di Troia},
  C.}, \au{{Fogaccia}, G.}, \au{{Vlad}, G.} \& \au{{Zonca}, F.}} \yr{2011}
  \at{An extended hybrid magnetohydrodynamics gyrokinetic model for numerical
  simulation of shear alfv{\'e}n waves in burning plasmas}.  \jt{Phys. Plasmas}
   \bvol{18}~(5),  \pg{052504}.

\bibitem[{Wang} {\em et~al.\/}(2012){Wang}, {Briguglio}, {Chen}, {Di Troia},
  {Fogaccia}, {Vlad} \& {Zonca}]{WB12pre}
{\sc \au{{Wang}, X.}, \au{{Briguglio}, S.}, \au{{Chen}, L.}, \au{{Di Troia},
  C.}, \au{{Fogaccia}, G.}, \au{{Vlad}, G.} \& \au{{Zonca}, F.}} \yr{2012}
  \at{Nonlinear dynamics of beta-induced alfv{\'e}n eigenmode driven by
  energetic particles}.  \jt{Phys. Rev. E}  \bvol{86}~(4),  \pg{045401(R)}.

\bibitem[{Wong} \& {Berk}(1998)]{WB98}
{\sc \au{{Wong}, H.V.} \& \au{{Berk}, H.L.}} \yr{1998}  \at{Growth and
  saturation of toroidal alfv{\'e}n eigenmode modes destabilized by ion
  cyclotron range of frequency produced tails}.  \jt{Phys. Plasmas}
  \bvol{5}~(7),  \pg{2781--2796}.

\bibitem[{Zaslavsky} {\em et~al.\/}(2008){Zaslavsky}, {Krafft}, {Gorbunov} \&
  {Volokitin}]{ZK08}
{\sc \au{{Zaslavsky}, A.}, \au{{Krafft}, C.}, \au{{Gorbunov}, L.} \&
  \au{{Volokitin}, A.}} \yr{2008}  \at{Wave-particle interaction at double
  resonance}.  \jt{Phys. Rev. E}  \bvol{77}~(5),  \pg{056407}.

\bibitem[{Zonca} {\em et~al.\/}(2015{\natexlab{{\em a\/}}}){Zonca}, {Chen},
  {Briguglio}, {Fogaccia}, {Milovanov}, {Qiu}, {Vlad} \& {Wang}]{ZC15ppcf}
{\sc \au{{Zonca}, F.}, \au{{Chen}, L.}, \au{{Briguglio}, S.}, \au{{Fogaccia},
  G.}, \au{{Milovanov}, A.V.}, \au{{Qiu}, Z.}, \au{{Vlad}, G.} \& \au{{Wang},
  X.}} \yr{2015{\natexlab{{\em a\/}}}}  \at{Energetic particles and multi-scale
  dynamics in fusion plasmas}.  \jt{Plasma Phys. Contr. Fusion}  \bvol{57}~(1),
   \pg{014024}.

\bibitem[{Zonca} {\em et~al.\/}(2015{\natexlab{{\em b\/}}}){Zonca}, {Chen},
  {Briguglio}, {Fogaccia}, {Vlad} \& {Wang}]{ZC15njp}
{\sc \au{{Zonca}, F.}, \au{{Chen}, L.}, \au{{Briguglio}, S.}, \au{{Fogaccia},
  G.}, \au{{Vlad}, G.} \& \au{{Wang}, X.}} \yr{2015{\natexlab{{\em b\/}}}}
  \at{Nonlinear dynamics of phase space zonal structures and energetic particle
  physics in fusion plasmas}.  \jt{New J. Phys.}  \bvol{17}~(1),  \pg{013052}.

\end{thebibliography}

\end{document}